\documentclass [10pt]{article}
\usepackage{amsmath,amssymb,cite}

\numberwithin{equation}{section}

\begin{document}
\begin{flushright}
UCD-2002-15\\
DIAS-02-06\\
\end{flushright}
\begin{center}
{\bf ON THE ORIGIN OF THE UV-IR MIXING IN NONCOMMUTATIVE MATRIX
GEOMETRY}
\bigskip

Sachindeo Vaidya$^\dagger$\footnote{vaidya@dirac.ucdavis.edu} and
Badis Ydri$^{*}$\footnote{ydri@synge.stp.dias.ie} \\
$^\dagger${\it Department of Physics, \\
University of California, Davis, CA 95616, USA.} \\
\bigskip
$^*${\it School of Theoretical Physics, \\
Dublin Institute for Advanced Studies, Dublin, Ireland.}\\

\end{center}

\begin{abstract}
Scalar field theories with quartic interaction are quantized on fuzzy
$S^2$ and fuzzy $S^2\times S^2$ to obtain the $2$- and $4$-point
correlation functions at one-loop. Different continuum limits of these
noncommutative matrix spheres are then taken to recover the quantum
noncommutative field theories on the noncommutative planes ${\mathbb
R}^2$ and ${\mathbb R}^4$ respectively. The canonical limit of large
stereographic projection leads to the usual theory on the
noncommutative plane with the well-known singular UV-IR mixing. A new
planar limit of the fuzzy sphere is defined in which the
noncommutativity parameter ${\theta}$, beside acting as a short
distance cut-off, acts also as a conventional cut-off
${\Lambda}=\frac{2}{\theta}$ in the momentum space. This
noncommutative theory is characterized by absence of UV-IR mixing. The
new scaling is implemented through the use of an intermediate scale
that demarcates the boundary between commutative and noncommutative
regimes of the scalar theory. We also comment on the continuum limit of the $4-$point function.
\end{abstract}

\section{Introduction and Results}

Noncommutative manifolds derive their interest not only from the fact
that they make their appearance in string theory (see for eg \cite{dn}
for a review of noncommutative geometry in string theory), but also
because they can potentially lead to natural ultra-violet
regularization of quantum field theories. The notion of
noncommutativity suggests a ``graininess'' for spacetime, and hence
can have interesting implications for models of quantum gravity.

Theoretical research has usually focused on either ``flat''
noncommutative spaces like ${\mathbb R}^{2n}$ or noncommutative tori
$T^{2n}$, or ``curved'' spaces that can be obtained as co-adjoint
orbits of Lie groups. In the latter category, attention has mostly
focused on using compact groups leading to noncommutative versions of
${\mathbb C}P^n$ \cite{bdlmo,abiy,trivai}, which are described by
finite-dimensional matrices and one or more size moduli: for example,
the fuzzy sphere is described by $N \times N$ matrices and its radius
$R$. (We use descriptions like ``flat'' or ``curved compact'' only in
a loose sense here).

Considerable attention has thus been devoted in trying to understand
properties of simple theories written on noncommutative manifolds. In
this endeavor, attention has most often been devoted to theories on
noncommutative ${\mathbb R}^{2n}$ and $ T^{2n}$ in the case of flat
spaces, and the curved space $S_F^2$ (the fuzzy sphere). Theories on
the noncommutative flat spaces generally possess infinite number of
degrees of freedom in contrast to those on ``compact'' spaces like
$S_F^2$. In either case, a key property of noncommutative theories
that is different from ordinary ones is the nature of the rule for
multiplying two functions. For example, the
star-product on ${\mathbb R}^{2n}$ (involving the noncommutativity
parameter $\theta$) is used for noncommutative theories, while
ordinary theories use the usual point-wise multiplication. On the
other hand, functions on curved compact noncommutative spaces are
simply finite-dimensional matrices, and are multiplied by the usual
matrix multiplication. This makes theories on ``curved''
noncommutative spaces easier to study numerically (although it must
also be mentioned here that the torus with rational noncommutativity
can also be studied using finite-dimensional matrices \cite{amns}).

Working on curved compact spaces also allows us to study the
flattening limit, which is when we take matrix size as well as the
length moduli to infinity. For example, the fuzzy sphere $S_F^2$ can
be flattened to give us the noncommutative plane. In this limit, we
expect to reproduce the behavior of the theory on the flat
manifold. Surprisingly, this limit can be crafted in a variety of
ways.

A simple way to understand this is as follows. All dimensionful
quantities can be expressed in terms of ``radius moduli'', i.e. the
length scale that defines the size of the compact space. Continuum
limit usually corresponds to taking the size of the matrices to
infinity, while flattening corresponds to taking large radii. However,
there is a large family of scales available to us in this flattening
limit. In other words, there are many ways of getting a relevant
length dimension quantity on the non-compact space. We could scale
both $R$ and $N$ to infinity keeping $R/N^\alpha$ fixed, where
$\alpha$ is some number. This corresponds to a length scale on the
plane, and all quantities in the quantum field theory (QFT) on the
plane can be measured with respect to this scale. A priori, one would
suspect that different values of $\alpha$ can lead to theories that
behave dramatically differently.

As we will argue here, this variety in the choice of scaling gives us
a refined probe to understand the nature of noncommutativity more
clearly. In particular, we will show with two different scaling limits
how this works. One corresponds to ``strongly'' noncommutative
theories, possessing singular properties like UV-IR mixing that makes
it impossible to write down corresponding low-energy Wilsonian
actions. The other corresponds to ``weak'' noncommutativity in a sense
that we will make precise. Briefly, these weakly noncommutative
theories are defined on a noncommutative plane, but do not exhibit
UV-IR mixing. In some sense, these theories mark the edge between
noncommutativity and commutativity.

The standard method of investigating perturbative properties of a
scalar QFT is by introducing an ultra-violet cut-off (see for example
\cite{wilson}). Instead of working with arbitrarily high energies, one
works with the partition function of this cut-off theory, and attempts
to study quantities that depend only weakly on the UV
cut-off. However, applying this technique to noncommutative theories
is problematic \cite{mirase}: taking the limit of small external
momentum does not commute with taking the limit of infinite
cut-off. This problem is commonly known as UV-IR mixing.

QFTs on noncommutative curved spaces allow us to implement a finer
version of the above procedure. In addition to the natural UV cut-off
(characterized by $1/N$ where $N$ is the matrix size), we can
introduce an intermediate scale $1/j$ characterized by an integer
$j<N$. It is the interplay between $j,N \rightarrow \infty$ and $R
\rightarrow \infty$ that we will exploit to understand the ``edge''
between commutativity and noncommutativity. 

In this article, we make concrete this set of ideas by applying them
to $S_F^2$ and $S_F^2 \times S_F^2$. The former is characterized by
$(2l+1) \times (2l+1)$ matrices and radius $R$, the latter by two
copies of the same matrix algebra and two radii $R_1$ and
$R_2$. Flattening these spaces by taking $l$ and $R_i$ to infinity (in
a prescribed manner) gives us noncommutative ${\mathbb R}^2$ and
${\mathbb R}^4$ respectively.

In particular, we will study two such scalings here. For example for
$S_F^2$, we keep $\theta' = R/\sqrt{l}$ fixed in the first case, and
keep $\theta = R/l$ fixed in the second, as we take $l$ and $R$ to
infinity. The former gives us the usual theory on the noncommutative
plane, which at the one-loop level reproduces the singularities of
UV-IR mixing. The latter is a new limit, and corresponds to keeping
the UV cut-off fixed in terms of the noncommutative parameter
$\theta$.

A short version explaining the new scaling limit appeared in
\cite{vayd}.

The fuzzy sphere is described by three matrices $x_i^F = \theta
L_i$ where $L_i$'s are the generators of $SU(2)$ for the spin $l$
representation and $\theta$ has dimension of length. The radius
$R$ of the sphere is related to $\theta$ and $l$ as $ R^2 =
\theta^2 l(l + 1)$. The usual action for a matrix model on
$S_F^2$ is
\begin{equation}
S = \frac{R^2}{2l+1} {\rm Tr}\; \left( \frac{[L_i, \Phi]^{\dagger}
[L_i , \Phi]}{R^2} + m^2 \Phi^2 + V[\Phi] \right),
\label{fs2action}
\end{equation}
and has the right continuum limit as $l \rightarrow \infty$. Because
of the noncommutative nature of $S_F^2$, there is a natural
ultra-violet (UV) cut-off: the maximum energy ${\Lambda}^2_{max}$ is
$=2l(2l+1)/{R^2}$. To get the theory on a noncommutative plane, the
usual strategy is to restrict to (say) the north pole, define the
noncommutative coordinates as $x_a^{NC}{\equiv}x_a^{F}$, ($a=1,2$),
and then take both $l$ and $R$ to infinity in a precisely specified
manner. For example, a commonly used limit requires us to hold
${\theta}^{'} = R/\sqrt{l}$ fixed as both $R$ and $l$ increase, which
gives us a noncommutative plane with \cite{perelomov,kishimoto}
\begin{equation} 
[{x}_1^{NC}, {x}_2^{NC}]=-i {\theta}^{'2}.
\label{limit1}
\end{equation} 
It is easy to see that in this limit, $\Lambda_{max}$ diverges, while
$\theta$ tends to zero. This is the analogous to the standard
stereographic projection.

A second scaling limit which is of interest to us here is one in which
$R$ and $l$ become large with noncommutativity parameter given now by
$\theta=R/l$ kept fixed. The above noncommutativity relation becomes
simply $[x_1^{NC},x_2^{NC}]=-iR\theta$ which means that $x_a^{NC}$'s
are now strongly noncommuting coordinates ( $R{\rightarrow}{\infty}$ )
and hence nonplanar amplitudes are expected to simply drop out in
accordance with \cite{mirase}. This can also be seen from the fact
that in this scaling (as is obvious from the relation
$R^2={\theta}^2l(l+1)$) $\Lambda_{max}$ no longer diverges: it is now
of order $1/\theta$, and there are no momentum modes in the theory
larger then this value. Alternatively we will also show that in this
limit the noncommutative coordinates can be instead identified as
$X_{a}^{NC}=x_a^{NC}/{\sqrt{l}}$ with noncommutative structure
\begin{equation} 
[{X}_1^{NC}, {X}_2^{NC}]=-i {\theta}^2.\label{limit2}
\end{equation} 

While this scaling for obtaining ${\mathbb R}^2_{\theta}$ is simply
stated , obtaining the corresponding theory with the above criteria is
somewhat subtle. Indeed passing from $[x_1^{NC},x_2^{NC}]=-iR\theta$
to (\ref{limit2}) corresponds in the quantum theory to a re-scaling of
momenta sending thus the finite cut-off $\Lambda=\frac{2}{\theta}$ to
infinity. In order to bring the cut-off back to a finite value
${\Lambda}_x=x{\Lambda}$, where $x$ is an arbitrary positive real
number, we modify the Laplacian on the fuzzy sphere
${\Delta}=[L_i,[L_i,..]]$ so that to project out modes of momentum
greater than a certain value $j$ given by
$j=[\frac{2\sqrt{l}}{x}]$. In other words, the theory on the
noncommutative plane ${\mathbb R}^2_{\theta}$ with UV cut-off
${\theta}^{-1}$ is obtained by flattening not the full theory on the
fuzzy sphere but only a ``low energy'' sector. One can argue that only
for when ${\Lambda}_x={\Lambda}$ that the canonical UV-IR
singularities become smoothen out. At this value we have
$j=[2\sqrt{l}]$ which marks somehow the boundary between commutative
and noncommutative field theories.

The generalization to noncommutative ${\mathbb R}^4$ is obvious. We
work on $S_F^2 \times S_F^2$ and then take the scaling limit with
$\theta$ fixed, which is the case of most interest in this article. By
analogy with (\ref{fs2action}), the scalar theory with quartic
self-interaction on $S_F^2 \times S_F^2$ is
\begin{eqnarray}
S =\frac{R_a^2}{2l_a +1}\frac{R_b^2}{2l_b +1}
   Tr_a Tr_b \left( \frac{[L_i^{(a)},
  \Phi]^{\dagger}[L_i^{(a)}, \Phi]}{R_a^2} + \frac{[L_i^{(b)},
  \Phi]^{\dagger}[L_i^{(b)},\Phi]}{R_b^2} + {\mu}_l^2 \Phi^2 +
  \frac{{\lambda}_4}{4!}\Phi^4 \right),
\label{s2s2action}
\end{eqnarray}
where $a$ and $b$ label the first and the second sphere respectively,
and $L_i^{(a,b)}$'s are the generators of rotation in spin
$l_{a,b}$-dimensional representation of $SU(2)$, and $\Phi$ is a
$(2l_a +1)\times (2l_a +1) \otimes (2l_b +1) \times (2l_b +1)$
hermitian matrix. As $l_a, l_b$ go to infinity, we recover the scalar
theory on an ordinary $S^2 \times S^2$.

Our strategy for obtaining the theory on noncommutative ${\mathbb
R}^4$ is straightforward: as discussed in \cite{gkp,vaidya}, we expand
$\Phi$ of action (\ref{s2s2action}) in terms of $SU(2)$ polarization
tensors (for definition and various properties of polarization
tensors, see for example \cite{VMK}). Using standard perturbation
theory and a conventional renormalization procedure, we calculate the
two- and four-point correlation functions, and then we scale $R,l
\rightarrow \infty$ with $\theta$ fixed.  Actually (and as we just
have said), implementation of the new scaling is somewhat subtle, in
that we will need to work not with the full theory on $S_F^2 \times
S_F^2$ but with a suitably defined low-energy sector. This low-energy
sector is selected by projecting out the high energy modes in an
appropriate manner using projection operators, and thus working with a
modified Laplacian:
\begin{equation} 
{\Delta}_j = {\Delta}+ \frac{1}{\epsilon}(1-P_j), \quad
j = [2\sqrt{l}],\nonumber
\end{equation}
where $P_j$ is the projector on all the modes associated with the
eigenvalues $k=0,...,j$, and ${\Delta}$ is the canonical Laplacian on
the full fuzzy sphere $S_F^2 \times S_F^2$. The flattening limit
(\ref{limit2}) is thus implemented on the scalar field theory
(\ref{s2s2action}) as the limit in which we first take
${\epsilon}{\rightarrow}0$ above, then we proceed with
$R,l{\rightarrow}{\infty}$ keeping ${\theta}=\frac{R}{l}$ fixed.

An obvious consequence of our scaling procedure is that the
correlation functions are not singular functions of external momenta.

There is a nice intuitive explanation for using the modified
Laplacian. If the momenta are cut-off at too low a value, the system
becomes in the commutative regime, while if the cut-off is too close
to $l$, the system remains noncommutative. The choice $[2\sqrt{l}]$
for the cut-off is in some sense the edge between these two
situations: there is some noncommutativity in the behaviour, but there
is no UV-IR mixing. We will have more to say about this in section 4.

The paper is organized as follows: in section $2$ we quantize
${\phi}^4$ theory on $S^2_F \times S^2_F$ and obtain the one-loop
corrections to the $2$-point and $4-$point functions . We also define in this section the precise meaning of UV-IR mixing
on ${\bf S}^2_F{\times}{\bf S}^2_F$ and write down the effective action . Section $3$ is the central importance, in which we
define continuum planar limits of the fuzzy sphere. In particular we
show how the singular UV-IR mixing emerges in the canonical limit of
large stereographic projection of the spheres onto planes. We also
show that in a new continuum flattening limit, a natural momentum
space cut-off (inversely proportional to the noncommutativity
parameter $\theta$) emerges, and as a consequence the UV-IR mixing is
completely absent. Section $3$ contains also the computation of the
continuum limit of the $4-$point function. As it turns out we recover exactly the planar
one-loop correction to the $4$-point function on noncommutative
${\mathbb R}^4$. We conclude in section $4$ with some general
observations.

\section{Effective Action on ${\bf S}^2_F{\times}{\bf S}^2_F$}
In this section, we will set up the quantum field theory on $S_F^2
\times S_F^2$, making explicit our notation and conventions. These
reflect our intent to consider $S_F^2 \times S_F^2$ as a discrete
approximation of noncommutative ${\mathbb R}^4$.

Each of the spheres $(\sum_i x_i^{(a)} x_i^{(a)} = {R^{(a)}}^2,
a=1,2)$ is approximated by the algebra $Mat_{2l_i +1}$ of $(2l_i +1)
\times (2l_i +1)$ matrices. The quantization prescription is given as
usual, by
\begin{equation}
n_i^{(a)} = \frac{x_i^{(a)}}{R^{(a)}} \quad {\rightarrow} \quad n_i^{(a)F} =
 \frac{L_i^{(a)}}{\sqrt{l_a(l_a+1)}}. 
\label{4}
\end{equation}
This prescription follows naturally from the canonical quantization of
the symplectic structure on the classical sphere (see for example
\cite{badisthesis}) by treating it as the co-adjoint orbit
$SU(2)/U(1)$. The $L_i^{(a)}$'s above are the generators of the IRR
representation $l_a$ of $SU(2)$: they satisfy $
[L_i^{(a)},L_j^{(a)}]=i{\epsilon}_{ijk}L_k^{(a)}$ and
$\sum_{i=1}^3L_i^{(a)2}=l_a(l_a+1)$. Thus
\begin{eqnarray}
[n_i^{(a)F},n_j^{(b)F}] = \frac{i}{\sqrt{l(l+1)}} {\delta}^{ab}
{\epsilon}_{ijk}n_{k}^{(a)F}.
\label{early}
\end{eqnarray}
Formally, $S_F^2 {\times} S_F^2$ is the algebra ${\bf
A}=Mat_{2l_1+1}{\otimes}Mat_{2l_2+1}$ generated by the identity ${\bf
1}{\otimes}{\bf 1}$ together with $L_i^{(1)}{\otimes}{\bf 1}$ and
${\bf 1}{\otimes}L_i^{(2)}$. This algebra ${\bf A}$ acts trivially on
the $(2l_1+1)(2l_2+1)$-dimensional Hilbert space ${\cal H}={\cal
H}_{1}{\otimes}{\cal H}_{1}$ with an obvious basis $\{|l_1m_1 \rangle
|l_2m_2 \rangle\}$.

The fuzzy analogue of the continuum derivations ${\cal
L}_i^{(a)}=-i{\epsilon}_{ijk}n_j^{(a)}{\partial}^{(a)}_k$ are given by
the adjoint action: we make the replacement
\begin{equation} 
{\cal L}_{i}^{(a)}{\rightarrow}K_{i}^{(a)} =
L_{i}^{(a)L}-L_{i}^{(a)R}.
\label{lapl}
\end{equation} 
The $L_{i}^{(a)L}$'s generate a left $SO(4)$ (more precisely
$SU(2){\otimes}SU(2)$) action on the algebra ${\bf A}$ given by
$L_i^{(a)L}M=L_i^{(a)}M$ where $M{\in}{\bf A}$. Similarly, the
$L_{i}^{(a)R}$'s generate a right action on the algebra, namely
$L_{i}^{(a)R}M=ML_{i}^{(a)}$. Remark that $K_{i}^{(a)}$'s annihilate
the identity ${\bf 1}{\otimes}{\bf 1}$ of the algebra ${\bf A}$ as is
required of a derivation.

In fact, it is enough to set $l_a = l_b = l$ and $R_a = R_b = R$ as
this corresponds in the limit to a noncommutative ${\mathbb R}^4$ with
a Euclidean metric on ${\mathbb R}^2{\times}{\mathbb R}^2$. The
general case simply corresponds to different deformation parameters in
the two ${\mathbb R}^2$ factors, and the extension of all results is
thus obvious (see equation $(6)$ of \cite{dn}).

In close analogy with the action on continuum $S^2{\times}S^2$, we put
together the above ingredients to write the action on $S_F^2 \times
S_F^2$:
\begin{equation}
{\cal S}_{l} = \frac{R^4}{(2l+1)^2}{\rm Tr}_{\cal H}\bigg[
  \frac{1}{R^2} \hat{\Phi}[L_i^{(1)},[L_i^{(1)},\hat{\Phi}]] +
  \frac{1}{R^2}\hat{\Phi} [L_i^{(2)},[L_i^{(2)},\hat{\Phi}]]
  +{\mu}_{l}^2{\hat{\Phi}}^2 + V(\hat{\Phi})\bigg] \equiv {\cal
  S}^{(0)}_l + {\cal S}^{int}_l.
\label{action}
\end{equation}
This action has the correct continuum (i.e. $l\rightarrow \infty, R$
fixed) limit:
\begin{equation} 
{\cal S}_\infty = R^4\int_{S^2} \frac{d{\Omega}^{(1)}}{4{\pi}}
\frac{d{\Omega}^{(2)}}{4{\pi}} \bigg[\frac{1}{R^2} {\Phi}{\cal
L}_{i}^{(1)}{\cal L}_{i}^{(1)} ({\Phi}) + \frac{1}{R^2}{\cal
L}_i^{(2)}{\cal L}_i^{(2)} ({\Phi}) + {\mu}_{\infty}^2{\Phi}^2+
V({\Phi})\bigg].
\end{equation} 
While the technology presented here can be applied to any polynomial
potential, we will restrict ourselves to $V(\hat\Phi) =
\frac{{\lambda}_{4,l}}{4!}{\hat{\Phi}}^4$. We have explicitly
introduced factors of $R$ wherever necessary to sharpen the analogy
with flat-space field theories: the integrand $R^4 d\Omega_1
d\Omega_2$ has canonical dimension of (Length)$^4$ like $d^4x$, the
field has dimension (Length)$^{(-1)}$, $\mu_l$ has (Length)$^{(-1)}$
and $\lambda_{4,l}$ is dimensionless.

Following \cite{gkp,vaidya}, the fuzzy field $\hat{\Phi}$ can
be expanded in terms of polarization operators \cite{VMK} as
\begin{eqnarray}
\hat{\Phi} &=& (2l+1)\sum_{k_1=0}^{2l}\sum_{m_1 =
  -k_1}^{k_1}\sum_{p_1=0}^{2l}\sum_{n_1=-p_1}^{p_1}
    {\phi}^{k_1m_1p_1n_1} T_{k_1m_1}(l){\otimes}T_{p_1n_1}(l)\nonumber\\
&\equiv&(2l+1)\sum_{11}{\phi}^{11}T_{1}(l_1){\otimes}T_{1}(l_2)
    \label{expansion} 
\end{eqnarray}
In our shorthand notation $\phi^{11}$ (for $\phi^{k_1m_1p_1n_1}$), the
quantum numbers from the first sphere come with subscript $1$ (as in
$(k_1 m_1)$, as do those for the second sphere.
 
The $T_{km}(l)$ are the polarization tensors which satisfy
\begin{eqnarray}
K_{\pm}^{(a)}T_{k_1m_1}(l) &=& {\mp}\frac{1}{\sqrt{2}}\sqrt{k_1(k_1+1)
  - m_1(m_1{\pm}1)}T_{k_1m_1{\pm}1}(l),\nonumber\\
K^{(a)}_3T_{k_1m_1}(l) &=& m_1T_{k_1m_1}(l),\nonumber\\
(\vec{K}^{(a)})^2T_{k_1m_1}(l) &=& k_1(k_1+1)T_{k_1m_1}(l),\nonumber
\end{eqnarray}
and the identities
\begin{eqnarray} 
{\rm Tr}_{\cal H}T_{k_1m_1}(l)T_{p_1n_1}(l) =
(-1)^{m_1}{\delta}_{k_1p_1}{\delta}_{m_1+n_1,0},\quad
T^{\dagger}_{k_1m_1}(l)=(-1)^{m_1}T_{k_1-m_1}(l).\nonumber 
\end{eqnarray} 
The field $\hat{\Phi}$ has a finite number of degrees of freedom,
totaling to $(2l_1 +1)^2 (2l_2 +2)^2$.

Our interest is restricted to hermitian fields since they are the
analog of real fields in the continuum. Imposing hermiticity
$\hat{\Phi}^{\dagger}=\hat{\Phi}$, we obtain the conditions
$\bar{\phi}^{k_1m_1p_1n_1} = (-1)^{m_1+n_1}{\phi}^{k_1-m_1p_1-n_1}$.

Since the field on our fuzzy space has only a finite number of degrees
of freedom, the simplest and most obvious route to quantization is via
path integrals. The partition function
\begin{equation} 
{\cal Z}=N\int {\cal D}{\phi} e^{-S_{l}-S^{int}_{l}}, \quad {\cal
D}{\phi}=\prod_{11}\frac{d\bar{\phi}^{11}d{\phi}^{11}}{2{\pi}}
\label{partfn}
\end{equation} 
for the theory yields the (free) propagator
\begin{equation} 
\langle {\phi}^{k_1m_1p_1n_1}{\phi}^{k_2m_2p_2n_2} \rangle =
\frac{(-1)^{m_2+n_2}}{R^2} \frac{{\delta}^{k_1k_2}{\delta}^{m_1,-m_2}
  {\delta}^{p_1p_2} {\delta}^{n_1,-n_2}}{k_1(k_1+1) + p_1(p_1+1) +
  R^2{\mu}^2_{l}}.
\label{propagator}
\end{equation} 
The Euclidean ``$4$-momentum'' in this setting is given by
$(11){\equiv}(k_1,m_1,p_1,n_1)$ with ``square''
$(11)^2=k_1(k_1+1)+p_1(p_1+1)$. For quartic interactions, the vertex
is given by the expression
\begin{equation} 
{\cal S}^{int}_{l} =
\sum_{11} \sum_{22} \sum_{33} \sum_{44} V(11,22,33,44)
    {\phi}^{11}{\phi}^{22} {\phi}^{33}{\phi}^{44},
\label{4-interaction}
\end{equation} 
with
\begin{eqnarray} 
V(11,22,33,44) &=& R^4 \frac{{\lambda}_4}{4!} V_1(1234,km)
V_2(1234,pn), \quad {\rm where} \nonumber \\
V_1(1234,km) &=& (2l+1){\rm Tr}_{H_{1}}
\bigg[T_{k_1m_1}(l)T_{k_2m_2}(l)T_{k_3m_3}(l)T_{k_4m_4}(l)\bigg]
\end{eqnarray} 
and similarly for $V_2(1234,pn)$.
\subsection{The $2-$Point Function}
The energy of each mode ${\phi}^{k_1m_1p_1n_1}$ is the square of the
fuzzy $4$-momentum, namely $(11)^2=k_1(k_1+1)+p_1(p_1+1)$. Since
$m_1=-k_1,\cdots,k_1$ and $n_1=-p_1,\cdots,p_1$, there are
$(2k_1+1)(2p_1+1)$ modes with the same energy for each pair of values
$(k_1,p_1)$, and may thus be thought of as naturally forming an energy
shell. Integrating out the high energy modes (with
$(k_1=2l_1,p_1=2l_2)$ in the path integral implements for us the
``shell'' approach to renormalization group adapted to fuzzy space
field theories \cite{vaidya}.

Integrating out only over the high momentum modes
$1_f1_f{\equiv}(k=2l_1,m,p=2l_2,n)$, the terms in the action that
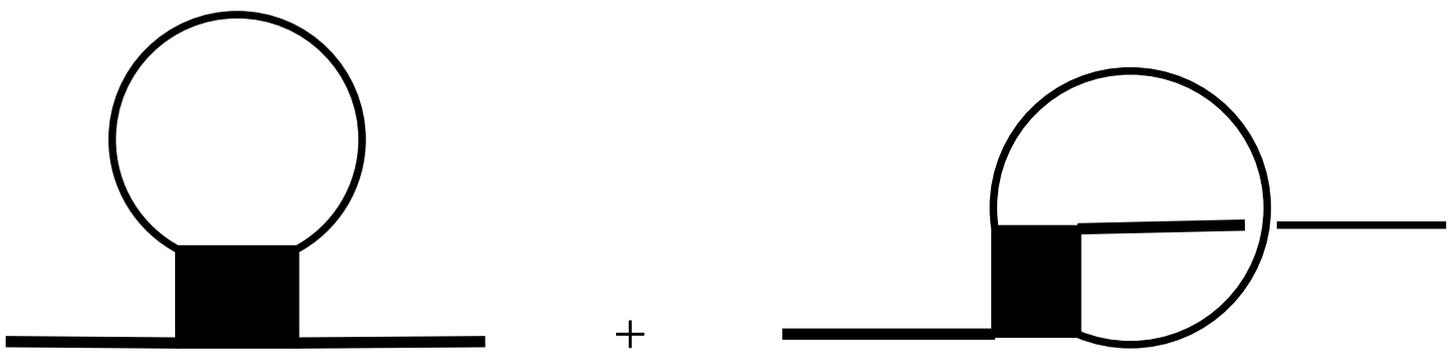
contribute to the $2$-point function at one-loop are given by
\begin{eqnarray}
{\Delta}{\cal S}_2^{(1)} &=& \ldots + 4\sum_{\bar{1}\bar{1}}
\sum_{\bar{2}\bar{2}} \sum_{3_f3_f} \sum_{4_f4_f}
V(\bar{1}\bar{1},\bar{2}\bar{2},3_f3_f,4_f4_f) {\phi}^{\bar{1}\bar{1}}
{\phi}^{\bar{2}\bar{2}} {\phi}^{3_f3_f} {\phi}^{4_f4_f}\nonumber\\ 
&+& 2 \sum_{\bar{1}\bar{1}} \sum_{{2_f2_f}} \sum_{\bar{3}\bar{3}}
\sum_{4_f4_f} V(\bar{1}\bar{1},{2_f2_f},\bar{3}\bar{3},4_f4_f)
{\phi}^{\bar{1}\bar{1}} {\phi}^{{2_f2_f}} {\phi}^{\bar{3}\bar{3}}
{\phi}^{4_f4_f}+ \ldots
\label{ibp}
\end{eqnarray}
The ellipsis indicate omitted terms that are unimportant for the
2-point function calculation. The notation is that of equations
(\ref{expansion}) and (\ref{4-interaction}), and
$\sum_{\bar{1}\bar{1}}=\sum_{11}-\sum_{1_f1_f}$. The relative factor
in the above is $4$ to $2$ since there are $4$ ways to contract two
neighboring fields (i.e. planar diagrams) and only two different ways
to contract non-neighboring fields (non-planar diagrams). The relevant graphs are displayed in figure $1$ .

Instead of integrating out only one shell, one could integrate out an
arbitrary number of them. For example, integrating out $q^2$ shells
gives
\begin{eqnarray}
{\Delta}{\cal S}_2^{(q^2)} &=& \ldots + 4\sum_{\hat{1}\hat{1}}
\sum_{\hat{2}\hat{2}} \sum_{3_f3_f} \sum_{4_f4_f}
V(\hat{1}\hat{1},\hat{2}\hat{2},3_f3_f,4_f4_f) {\phi}^{\hat{1}\hat{1}}
{\phi}^{\hat{2}\hat{2}} {\phi}^{3_f3_f}{\phi}^{4_f4_f}\nonumber\\
&+& 2 \sum_{\hat{1}\hat{1}} \sum_{{2_f2_f}} \sum_{\hat{3}\hat{3}}
\sum_{4_f4_f} V(\hat{1}\hat{1},{2_f2_f},\hat{3}\hat{3},4_f4_f)
{\phi}^{\hat{1}\hat{1}} {\phi}^{{2_f2_f}} {\phi}^{\hat{3}\hat{3}}
{\phi}^{4_f4_f}+\ldots
\label{q2shells}
\end{eqnarray}
with now
\begin{eqnarray} 
\sum_{1_f1_f} = \sum_{k=2l_1-(q-1)}^{2l_1}
\sum_{m=-k}^k\sum_{p=2l_2-(q-1)}^{2l_2}\sum_{n=-p}^p, \quad
\sum_{\hat{1}\hat{1}} = \sum_{11}-\sum_{1_f1_f}\nonumber
\end{eqnarray} 
while the partition function (\ref{partfn}) takes the form 
\begin{equation} 
{\cal Z} = \hat{N} \int {\cal D}\hat{\phi} e^{-\hat{\cal S}_l -
\hat{\cal S}_i^{int} - \langle \Delta {{\cal S}_2}^{q^2} \rangle_f +
\frac{\langle \Delta {{\cal S}_2^{q^2}}^2 \rangle_f - \langle \Delta
{\cal S}_2^{q^2} \rangle^2_f}{2} + \cdots}, \quad {\rm where} \;\;
{\cal D}\hat{\phi} = \prod_{\hat{1}\hat{1}}\frac{d
\bar{\phi}^{\hat{1}\hat{1}} d \phi^{\hat{1}\hat{1}}}{2 \pi}
\label{zq2}
\end{equation} 

For $l_1=l_2=l$, the full one-loop corresponds to integrating over all
shells i.e. $q=2l+1$. The corresponding effective action is

\begin{eqnarray}
<{\Delta}{\cal S}_2^{(q^2)}>|_{q=2l+1} &=& \frac{1}{R^2}\frac{{\lambda}_{4,l}}{4!} \sum_{k_1m_1p_1n_1}\bigg[{\delta}{\mu}_l^{P}+{\delta}{\mu}_l^{NP}(k_1,p_1)\bigg] 
|{\phi}^{k_1m_1p_1n_1}|^2 .
\label{q2shells1}
\end{eqnarray}
The 2-point function computation readily gives
us the renormalized mass:
\begin{equation} 
{\mu}_l^{2}(k_1,p_1) = {\mu}_l^{2} +
\frac{1}{R^2}\frac{{\lambda}_{4,l}}{4!} \bigg[{\delta}{\mu}_l^{P} +
{\delta}{\mu}_l^{NP}(k_1,p_1)\bigg] 
\end{equation} 
where the planar contribution given by
\begin{equation} 
{\delta}{\mu}_l^{P} = 4\sum_{a=0}^{2l} \sum_{b=0}^{2l}A(a,b), \quad
A(a,b) =\frac{(2a+1)(2b+1)}{a(a+1)+b(b+1)+R^2{\mu}_l^2}.
\label{planar1}
\end{equation} 
On the other hand, the non-planar contribution is 
\begin{eqnarray} 
{\delta}{\mu}_l^{NP}(k_1,p_1) &=& 2 \sum_{a=0}^{2l} \sum_{b=0}^{2l}
A(a,b)(-1)^{k_1+p_1+a+b} B_{k_1p_1}(a,b), \quad {\rm where} \nonumber\\
B_{ab}(c,d)&=&(2l+1)^2 
\left\{ \begin{array}{ccc}
a&l   & l\\
c& l&l
\end{array}
\right\} \left\{
\begin{array}{ccc}
b&l  & l\\
d& l&l
\end{array}\right\}.
\label{nonplanar1}
\end{eqnarray}
The symbol $\{ \}$ in $B_{ab}(c,d)$ is the standard $6j$ symbol (see
for example \cite{VMK}). As is immediately obvious from these
expressions, both planar and non-planar graphs are finite and
well-defined for all finite values of $l$. However, a measure for the
fuzzy UV-IR mixing is the difference ${\Delta}$ between planar and
non-planar contributions, which we define below:
\begin{eqnarray}
{\delta}{\mu}_l^{P} + {\delta}{\mu}_l^{NP}(k_1,p_1) &=&
{\Delta}{\mu}_l^P + \frac{1}{2}{\Delta}(k_1,p_1), \quad
{\Delta}{\mu}_l^P = 6\sum_{a=0}^{2l}\sum_{b=0}^{2l}A(a,b) \nonumber\\
{\Delta}(k_1,p_1) &=& 4\sum_{a=0}^{2l} \sum_{b=0}^{2l} A(a,b)
\bigg[(-1)^{k_1+p_1+a+b} B_{k_1p_1}(a,b)-1\bigg].\label{regularized}
\end{eqnarray}
Were this difference $\Delta(k_1,p_1)$ to vanish, we would recover the
usual contribution to the mass renormalization as expected in a
commutative field theory. The fact that this difference is not zero in
the limit of large IRR's $l$, i.e.  $l{\rightarrow}{\infty}$, is what
is meant by UV-IR mixing on fuzzy ${\bf S}^2{\times}{\bf S}^2$. Indeed
this may be taken as the definition of the UV-IR problem on general
fuzzy spaces. In fact (\ref{regularized})
can also be taken as the regularized form of the UV-IR mixing on
${\bf R}^4$  . Removing the UV cut-off
$l{\longrightarrow}{\infty}$ while keeping the infrared cut-off
$R$ fixed $=1$ one can show that ${\Delta}$ diverges as $l^2$ ,
i.e
\begin{eqnarray}
{\Delta}(k_1,p_1)&{\longrightarrow}&(8l^2)\int_{-1}^{1}\int_{-1}^{1}\frac{dt_xdt_y}{2-t_x-t_y}\bigg[P_{k_1}(t_x)P_{p_1}(t_y)-1\bigg],\label{estimation1}
\end{eqnarray}
where , for simplicity , we have assumed ${\mu}_l<<l$
\cite{madore}. (\ref{estimation1}) is worse than the case of two
dimensions [ see equation $(3.20)$ of \cite{madore} ] , in here
not only the difference survives the limit but also it diverges .
This means in particular that the UV-IR mixing can be largely
controlled or perhaps understood if one understands the role of the
UV cut-off $l$ in the scaling limit and its relation to the
underlying star product on ${\bf S}^2_F$.
\subsection{The $4-$Point Function}
The computation of higher order correlation functions become very
complicated, but this exercise is necessary if we want to compute for example the
beta-function. It is also useful to put forward key features which
will be needed (in the future) to study noncommutative matrix gauge
theories and their continuum limits. We will only look at the
four-point function here.

Our starting point is (\ref{zq2}), which tells us that integrating out
$q^2$ shells produces the following correction to the $4-$point
function:
\begin{eqnarray}
\langle ({\Delta}S_2^{(q^2)})^2 \rangle_f &=&
W(\hat{1},\hat{2},\hat{3},\hat{5};\hat{4},\hat{6},\hat{7},\hat{8})
{\phi}^{1235} \bigg[\langle {\phi}^{4_f4_f} {\phi}^{6_f6_f} \rangle_f
\langle {\phi}^{7_f7_f}{\phi}^{8_f8_f} \rangle_f \nonumber\\
&+& \langle{\phi}^{4_f4_f} {\phi}^{7_f7_f} \rangle_f
\langle{\phi}^{6_f6_f} {\phi}^{8_f8_f} \rangle_f +
\langle{\phi}^{4_f4_f} {\phi}^{8_f8_f} \rangle_f
\langle{\phi}^{7_f7_f} {\phi}^{6_f6_f} \rangle_f\bigg],
\nonumber
\end{eqnarray}
Here, $W(\hat{1}, \hat{2}, \hat{3}, \hat{5}; \hat{4}, \hat{6},
\hat{7}, \hat{8}) = W_2(\hat{1} \hat{1}, \hat{2} \hat{2}, 4_f4_f,
6_f6_f) W_2(\hat{3} \hat{3}, \hat{5} \hat{5},7_f7_f, 8_f8_f)$ such
that $W_2(\hat{1} \hat{1}, \hat{2} \hat{2}, 3_f3_f, 4_f4_f) =
4V(\hat{1} \hat{1}, \hat{2} \hat{2}, 3_f3_f, 4_f4_f) + 2V(\hat{1}
\hat{1}, 3_f3_f, \hat{2}\hat{2},4_f4_f)$, ${\phi}^{1235} =
{\phi}^{\hat{1} \hat{1}} {\phi}^{\hat{2}\hat{2}} {\phi}^{\hat{3}
\hat{3}} {\phi}^{\hat{5}\hat{5}}$, and the notation is that of
equations (\ref{expansion}), (\ref{4-interaction}) and
(\ref{q2shells}). Inserting the free propagator (\ref{propagator})
above yields the $4$-point function
\begin{equation} 
\frac{\langle({\Delta} S_2^{(q^2)})^2 \rangle_f -
\langle{\Delta}S_2^{(q^2)} \rangle_f^2}{2} = \sum_{\hat{1}\hat{1}}
\sum_{\hat{2}\hat{2}} \sum_{\hat{3}\hat{3}} \sum_{\hat{5}\hat{5}} R^4
\frac{{\lambda}_4}{4!} {\phi}^{\hat{1}\hat{1}} {\phi}^{\hat{2}\hat{2}}
{\phi}^{\hat{3}\hat{3}} {\phi}^{\hat{5}\hat{5}}
{\delta}{\lambda}_4(1235)
\label{1000}
\end{equation} 
where
\begin{eqnarray}
{\delta}{\lambda}_4(1235) &=& \frac{{\lambda}_4}{4!} \sum_{k_4,k_6=f}
\sum_{p_4,p_6=f} \frac{A(k_4,p_4)
A(k_6,p_6)}{(2k_4+1)(2k_6+1)(2p_4+1)(2p_6+1)} \bigg[8{\eta}_1^{(1)}
{\eta}_2^{(1)} \nonumber\\
&+&16{\eta}_1^{(2)} {\eta}_2^{(2)} + 4{\eta}_1^{(3)} {\eta}_2^{(3)} +
8{\eta}_1^{(4)}{\eta}_2^{(4)}\bigg].
\label{1001}
\end{eqnarray}
The first graph in (\ref{1001}) is the usual one-loop contribution to
the $4-$point function , i.e the two vertices are planar. The fourth
graph contains also two planar vertices but with the exception that
one of these vertices is twisted , i.e with an extra phase. The second
graph contains on the other hand one planar vertex and one non-planar
vertex, whereas the two vertices in the third graph are both
non-planar. The relevant graphs are displayed in figures $2$ and $3$ . The analytic expressions for ${\eta}_i^{(a)} {\equiv}
{\eta}_i^{(a)}(k_4k_6;1235) = \sum_{m_4=-k_4}^{k_4} \sum_{m_6 =
-k_6}^{k_6}{\rho}_i^{(a)}(k_4k_6;1235)$ are given by
\begin{eqnarray}
{\rho}_i^{(1)} &=& (-1)^{m_4 + m_6} V_i(\hat{1} \hat{2}4_f6_f)
V_i(\hat{3} \hat{5}-4_f-6_f), \quad {\rho}_i^{(2)} = (-1)^{m_4 + m_6}
V_i(\hat{1} \hat{2}4_f6_f) V_i(\hat{3}-4_f\hat{5}-6_f) \nonumber\\
{\rho}_i^{(3)} &=& (-1)^{m_4 + m_6} V_i(\hat{1}4_f \hat{2}6_f)
V_i(\hat{3}-4_f \hat{5}-6_f), \quad {\rho}_i^{(4)} = (-1)^{m_4 + m_6}
V_i(\hat{1} \hat{2}4_f6_f) V_i(\hat{3} \hat{5}-6_f-4_f),\nonumber
\end{eqnarray}
where the lower index in $\eta$'s and $\rho$'s labels the sphere
whereas the upper index denotes the graph, and the notation $-4_f4_f$
stands for $(k_4,-m_4,p_4,-n_4)$ in contrast with
$4_f4_f = (k_4,m_4,p_4,n_4)$.

By using extensively the different identities in \cite{VMK} we can
find after a long calculation that the above $4$-point function has
the form
\begin{eqnarray}
{\delta}{\lambda}_4(1235) &=& \frac{{\lambda}_4}{4!} \bigg[ 8 {\delta}
{\lambda}_4^{(1)}(1235) + 16 {\delta}{\lambda}_4^{(2)}(1235) + 4
{\delta}{\lambda}_4^{(3)}(1235) + 8 {\delta}{\lambda}_4^{(4)}(1235)
\bigg], \quad \rm{where} \nonumber\\
{\delta}{\lambda}_4^{(a)}(1235) &=& \sum_{k_4,k_6=f} \sum_{p_4,p_6=f}
A(k_4,p_4) A(k_6,p_6) {\nu}^{(a)}_1(k_4k_6;1235)
{\nu}^{(a)}_2(p_4p_6;1235), \;\;a=1 \ldots 4,
\label{1002}
\end{eqnarray}
The label $f$ stands for the shells we integrated over and hence it
corresponds to $q^2=(2l+1)^2$ for the full one-loop contribution. The
planar amplitudes, in the first ${\mathbb R}^2$ factor for example,
are given by
\begin{equation} 
{\nu}_1^{(1)} = \sum_{k}(-1)^{k + k_4 + k_6} {\delta}_{k}(1235)
E_{k_1k_2}^{k_4k_6}(k) E_{k_3k_5}^{k_4k_6}(k), \quad {\nu}_1^{(4)} =
\sum_{k}{\delta}_{k}(1235) E_{k_1k_2}^{k_4k_6}(k)
E_{k_3k_5}^{k_4k_6}(k)
\label{planar4}
\end{equation} 
whereas the non-planar amplitudes are given by
\begin{equation} 
{\nu}_1^{(2)} = \sum_{k}(-1)^{k_3 + k_4} {\delta}_{k}(1235)
E_{k_1k_2}^{k_4k_6}(k) F_{k_3k_5}^{k_4k_6}(k), \quad {\nu}_1^{(3)} =
\sum_{k}(-1)^{k_2 + k_3}{\delta}_{k}(1235) F_{k_1k_2}^{k_6k_4}(k)
F_{k_3k_5}^{k_4k_6}(k)
\label{nonplanar4}
\end{equation} 
with
\begin{eqnarray}
F_{k_1k_2}^{k_4k_6}(k)&=&(2l+1)\sqrt{(2k_1+1)(2k_2+1)} \left\{
\begin{array}{ccc}
k_4&l  & l\\
k_6&l&l\\
k&k_1&k_2
\end{array}
\right\}\nonumber\\
E_{k_1k_2}^{k_4k_6}(k)&=&(2l+1)\sqrt{(2k_1+1)(2k_2+1)}\left\{
\begin{array}{ccc}
k_1&k_2  & k\\
l&l&l
\end{array}
\right\} \left\{
\begin{array}{ccc}
k_4&k_6  & k\\
l&l&l
\end{array}
\right\}.
\label{F}
\end{eqnarray}
The ``fuzzy delta'' function $\delta_k(1235)$ is defined by
\begin{equation} 
{\delta}_{k}(1235) = (-1)^m C^{km}_{k_1m_1k_2m_2}
C^{k-m}_{k_3m_3k_5m_5}.
\label{delta}
\end{equation} 
The justification for this name will follow shortly.

The full effective action at one-loop of the above scalar field theory on ${\bf S}^2_F{\times}{\bf S}^2_F$ is obtained by adding the two quantum actions (\ref{q2shells1}) and (\ref{1000}) to the classical action (\ref{action}) .

\section{Continuum Planar Limits}

We can now state with some detail the continuum limits in which the
fuzzy spheres approach (in a precise sense) the noncommutative
planes. There are primarily two limits of interest to us: one is the
canonical large stereographic projection of the spheres onto planes,
while the second is a new flattening limit which we will argue
corresponds to a conventional cut-off.

For simplicity, consider a single fuzzy sphere with cut-off $l$ and
radius $R$, and define the fuzzy coordinates $x_i^{F}={\theta}L_i^{}$
(i.e.  $x_{\pm}^{F}=x_1^{F}{\pm}ix_2^{F}$) where
${\theta}=R/{\sqrt{l(l+1)}}$. The stereographic projection
\cite{alpist,kishimoto} to the noncommutative plane is realized as
\begin{equation} 
y_{+}^{F}=2R^{}x_{+}^{F}\frac{1}{R-x_3^{F}}, \quad
y_{-}^{F}=2R\frac{1}{R-x_3^{F}}x_{-}^{F}. 
\end{equation} 
In the large $l$ limit it is obvious that these fuzzy coordinates
indeed approach the canonical stereographic coordinates. A planar
limit can be defined from above as follows:
\begin{equation} 
{\theta}^{'2}=\frac{{R}^{2}}{\sqrt{l(l+1)}}=\text{fixed as}\quad 
l,R{\rightarrow}{\infty}.
\label{contrast}
\end{equation} 
In this limit, the commutation relation becomes
\begin{equation} 
[y_{+}^{NC},y_{-}^{NC}] = -2{\theta}^{'2}, \quad
y_{\pm}^{NC}{\equiv}y_{\pm}^{F} = x_{\pm}^F,
\end{equation} 
where we have substituted $L_{3}=-l$ corresponding to the north
pole. The above commutation relation may also be put in the form
\begin{equation} 
[x^{NC}_1,x^{NC}_2] = -i{\theta}^{'2}, \quad x^{NC}_a{\equiv}x^{F}_a,
\quad a=1,2
\label{comm1}
\end{equation} 
The minus sign is simply due to our convention for the coherent states
on co-adjoint orbits. The extension to the case of two fuzzy spheres
is trivial.

A second way to obtain the noncommutative plane is by taking the limit
\begin{equation} 
{\theta}=\frac{R}{\sqrt{l(l+1)}}=\quad l,R{\rightarrow}{\infty}.
\label{flattening2}
\end{equation}
A UV cut-off is automatically built into this limit: the maximum
energy a scalar mode can have on the fuzzy sphere is $2l(2l+1)/R^2$,
which in this scaling limit is $4/\theta^2$. There are no modes with
energy larger than this value. To understand this limit a little
better, let us restrict ourselves to the north pole
$\vec{n}=\vec{n}_0=(0,0,1)$ where we have $\langle
\vec{n}_0,l|L_3|\vec{n}_0,l \rangle = -l$ and $\langle
\vec{n}_0,l|L_a|\vec{n}_0,l \rangle=0$, $a=1,2$. The commutator
$[L_1,L_2]=iL_3 = -il$, so the noncommutative coordinates on this
noncommutative plane ``tangential to the north pole'' can be given
either simply by $x_a^F$ as above. This now defines a strongly
noncommuting plane, viz
\begin{eqnarray}
[x_a^F,x_b^F]=-il{\theta}^2{\epsilon}_{ab}.\label{or}
\end{eqnarray}
Or aletrnatively one can define the noncommutative coordinate by 
${X}^{NC}_a{\equiv}\sqrt{\frac{\theta}{R}}x_a^F$, satisfying
\begin{equation}
[X^{NC}_a,X^{NC}_b]=-i{\theta}^2{\epsilon}_{ab}.
\label{commutationrelation}
\end{equation}
In the convention used here, ${\epsilon}_{12}=1$ and
${\epsilon}_{ac}{\epsilon}_{cb} = -{\delta}_{ab}$.

Intuitively, the second scaling limit may be understood as
follows. Noncommutativity introduces a short distance cut-off of the
order ${\delta}X=\sqrt{\frac{{\theta}^2}{2}}$ because of the
uncertainty relation $ {\Delta}X^{NC}_1 {\Delta}X^{NC}_2 {\geq}
\frac{{\theta}^2}{2}$. However, the Laplacian operators on generic
noncommutative planes do not reflect this short distance cut-off, as
they are generally taken to be the same as the commutative
Laplacians. On the above noncommutative plane
(\ref{commutationrelation}) the cut-off ${\delta}X$ effectively
translates into the momentum space as some cut-off ${\delta}P =
\frac{1}{\sqrt{2{\theta}^2}}$. This is because of (and in accordance
with) the commutation relations $[X^{NC}_a, P^{NC}_b] = i
\delta_{ab}$,$P_a^{NC} =
-\frac{1}{{\theta}^2}{\epsilon}_{ab}X_b^{NC}$, giving us the
uncertainty relations ${\Delta}X^{NC}_a {\Delta}P^{NC}_b {\geq}
\frac{{\delta}_{ab}}{2}$. Since one can not probe distances less than
${\delta}X$, energies above ${\delta}P$ should not be accessible
either, i.e.  $[P_a^{NC},P_b^{NC}] =
-\frac{i}{{\theta}^2}{\epsilon}_{ab}$. The fact that the maximum
energy of a mode is of order $1/\theta$ in the second scaling limit
ties in nicely with this expectation.

The limit (\ref{flattening2}) may thus be thought of as a
regularization prescription of the noncommutative plane which takes
into account our expectation of ``UV-finiteness'' of noncommutative
quantum field theories.

\subsection{Field Theory in the Canonical Planar Limit}

We are now in a position to study what happens to the scalar field
theory in the limit (\ref{comm1}). First we match the spectrum of the
Laplacian operator on each sphere with the spectrum of the Laplacian
operator on the limiting noncommutative plane as follows
\begin{equation} 
a(a+1)=R^2p_{a}^2,
\label{matchingcond0}
\end{equation} 
where $p_a$ is of course the modulus of the two dimensional momentum
on the noncommutative plane which corresponds to the integer $a$, and
has the correct mass dimension. However since the range of $a$'s is
from $0$ to $2l$, the range of $p_a^2$ will be from $0$ to $
\frac{2l(2l+1)}{R^2}=l{\Lambda}^{'2}{\rightarrow}\infty$,
${\Lambda}^{'}=2/{\theta}^{'}$. In other words, all information about
the UV cut-off is lost in this limit.

Let us see how the other operators in the theory scales in the above
planar limit. It is not difficult to show that the free action scales
as
\begin{equation} 
\sum_{a,b} \sum_{m_a,m_b} \bigg[R^2a(a+1)+R^2b(b+1)+R^4{\mu}_l^2\bigg]
|{\phi}^{abm_am_b}|^2 {\simeq} \int_{\sqrt{l}{\Lambda}^{'}}
\frac{d^2\vec{p}_ad^2\vec{p}_b}{{\pi}^2} \bigg[p_a^2+p_b^2+M^2\bigg]
|{\phi}_{NC}^{p_ap_b{\phi}_a{\phi}_b}|^2.
\label{dimensionanalysis0}
\end{equation} 
The scalar field is assumed to have the scaling property
${\phi}^{p_ap_b{\phi}_a{\phi}_b}_{NC}{\simeq} R^4{\phi}^{abm_am_b}$,
which gives the momentum-space scalar field the correct mass dimension
of $-3$ [recall that $[{\phi}^{abm_am_b}]=M$]. The ${\phi}_a$ and
${\phi}_b$ above (not to be confused with the scalar field!) are the
angles of the two momenta $\vec{p}_a$ and $\vec{p}_b$ respectively,
i.e. ${\phi}_a = \frac{{\pi}m_a}{a+\frac{1}{2}}$ and ${\phi}_b =
\frac{{\pi}m_b}{b+\frac{1}{2}}$. This formula is exact, and can be
simplified further when quantum numbers $a$'s and $b$'s are large: the
${\phi}_a$ and ${\phi}_b$ will be in the range $[-{\pi},{\pi}]$. It is
also worth pointing out that the mass parameter $M$ of the planar
theory is exactly equal to that on the fuzzy spheres,
i.e. $M={\mu}_{\infty}$, and no scaling is required. This is in
contrast with \cite{madore} but only due to our definition of the
fuzzy action (\ref{action}).

With these ingredients, it is not then difficult to see that the
flattening limit of the planar $2-$point function (\ref{planar1})
is given by
\begin{equation} 
\delta M^P {\equiv}\frac{{\delta}{\mu}_l^P}{R^2}= 16 \int \int
\frac{p_ap_bdp_adp_b}{p_a^2+p_b^2 + M^2}
\end{equation} 
which is the $2$-point function on noncommutative ${\mathbb R}^4$ with
a Euclidean metric ${\mathbb R}^2 \times {\mathbb R}^2$. By rotational
invariance it may be rewritten as
\begin{equation} 
{\delta} M^P = \frac{4}{{\pi}^2} \int_{\sqrt{l}{\Lambda}^{'}} \frac{d^4k}{k^2+M^2}.
\label{planar10}
\end{equation} 
We do now the same exercise for the non-planar $2$-point function
(\ref{nonplanar1}). Since the external momenta $k_1$ and $p_1$ are
generally very small compared to $l$ , one can use the following
approximation for the $6j$-symbols \cite{VMK}
\begin{eqnarray} 
\left\{\begin{array}{ccc}
           a&l  & l\\
           b& l&l
       \end{array} \right\}{\approx} \frac{(-1)^{a+b}}{2l} P_{a}
           (1-\frac{b^2}{2l^2}), \quad
           l{\rightarrow}{\infty},\;\;a<<l,\;0 {\leq} b{\leq}2l,
\label{appr1}
\end{eqnarray} 
By putting in all the ingredients of the planar limit we obtain the
result
\begin{eqnarray} 
{\delta}M^{NP}(k_1,p_1){\equiv}\frac{{\delta}{\mu}_l^{NP}}{R^2} = 8
\int_{0}^{\infty} \int_{0}^{\infty} 
\frac{p_ap_bdp_adp_b}{p_a^2 + p_b^2 + M^2} P_{k_1}(1 -
\frac{{\theta'}^4 p_a^2}{2R^2}) P_{p_1}(1 -
\frac{{\theta'}^4 p_b^2}{2R^2}). \nonumber
\end{eqnarray} 
Although the quantum numbers $k_1$ and $p_1$ in this limit are very
small compared to $l$, they are large themselves
i.e. $1<<k_1,p_1<<l$. On the other hand, the angles ${\nu}_{a}$
defined by $\cos{\nu}_{a}=1-\frac{{\theta'}^4 p_a^2}{2R^2}$ can be
considered for all practical purposes small, i.e. ${\nu}_a =
\frac{{\theta'}^2 p_a}{R}$ because of the large $R$ factor, and hence
we can use the formula (see for eg \cite{magnus}, page $72$)
\begin{eqnarray} 
P_{n}(\cos{\nu}_a) = J_{0}(\eta) + \sin^2\frac{{\nu}_a}{2}
\bigg[\frac{J_{1}(\eta)}{2{\eta}} - J_{2}(\eta) + \frac{\eta}{6}
J_{3}(\eta)\bigg] + O(\sin^4\frac{{\nu}_a}{2}),
\label{formula}
\end{eqnarray} 
for $n>>1$ and small angles ${\nu}_a$, with $\eta = (2n+1) \sin
\frac{{\nu}_a}{2}$. To leading order we then have
\begin{eqnarray} 
P_{k_1}(1 - \frac{{\theta'}^4 p_a^2}{2R^2}) =
J_0({\theta'}^2 p_{k_1}p_a) = \frac{1}{2{\pi}} \int_{0}^{2{\pi}}
d{\phi}_ae^{i{\theta'}^2 \cos{\phi}_a p_{k_1}p_a}.\nonumber
\end{eqnarray}
This result becomes exact in the strict limit of $l,R \rightarrow
\infty$ where all fuzzy quantum numbers diverge with ${R}$. We get
then
\begin{eqnarray} 
\delta M^{NP}(p_{k_1},p_{p_1}) = \frac{2}{{\pi}^2} \int
\int \int \int
\frac{(p_adp_ad{\phi}_a) (p_bdp_bd{\phi}_b)}{p_a^2+p_b^2 +
M^2}e^{i{\theta'}^2 p_{k_1}(p_a
\cos{\phi}_a)}e^{i{\theta'}^2 p_{p_1}(p_b \cos{\phi}_b)}.\nonumber
\end{eqnarray} 
By rotational invariance we can set ${\theta'}^2 B^{{\mu}{\nu}}
p_{k_1\mu} p_{a\nu} = {\theta'}^2 p_{k_1} (p_a \cos{\phi}_a)$, where
$B^{12}=-1$. In other words, we can always choose the two-dimensional
momentum $p_{k_1}$ to lie in the $y$-direction, thus making ${\phi}_a$
the angle between $\vec{p}_a$ and the $x$-axis. The same is also true
for the other exponential. We thus obtain the canonical non-planar
$2$-point function on the noncommutative ${\mathbb R}^4$ (with
Euclidean metric ${\mathbb R}^2 \times {\mathbb R}^2$). Again by
rotational invariance, this non-planar contribution to the $2$-point
function may be put in the compact form
\begin{eqnarray} 
\delta M^{NP}(p) = \frac{2}{{\pi}^2}\int_{\sqrt{l}{\Lambda}^{'}} \frac{d^4k}{k^2 + M^2}
e^{i{\theta'}^2 p B k}.
\label{nonplanar10}
\end{eqnarray}
The structure of the effective action in momentum space allows us to
deduce the star products on the underlying noncommutative space. For
example, by using the tree level action (\ref{dimensionanalysis0})
together with the one-loop contributions (\ref{planar10}) and
(\ref{nonplanar10}) one can find that the effective action obtained in
the large stereographic limit (\ref{contrast}) is given by
\begin{eqnarray} 
\int_{\sqrt{l}{\Lambda}^{'}} \frac{d^4\vec{p}}{(2{\pi})^4}\frac{1}{2} \bigg[\vec{p}^2 + M^2 +
\frac{g_4^2}{6} \big[2\int_{\sqrt{l}{\Lambda}^{'}} \frac{d^4k}{(2{\pi})^4} \frac{1}{k^2 + M^2}
+ \int_{\sqrt{l}{\Lambda}^{'}} \frac{d^4k}{(2{\pi})^4} \frac{e^{i{\theta}^{'2} \vec{p}B
\vec{k}}}{k^2 + M^2}\big]\bigg]|{\phi}_{1}(\vec{p})|^2
\label{effective0}
\end{eqnarray} 
where $g^2_4=8{\pi}^2{\lambda}_4$ and ${\phi}_{1}(\vec{p}) =
4{\pi}\sqrt{2}{\phi}_{NC}^{p_ap_b{\phi}_a{\phi}_b}$ and
$\sqrt{l}{\Lambda}^{'}{\rightarrow}{\infty}$. This effective 
action can be obtained from the quantization of the action
\begin{eqnarray} 
\int d^4x \bigg[\frac{1}{2}({\partial}_{\mu}{\phi}_1)^2 +
\frac{1}{2}M^2{\phi}_1^2 + \frac{g_4^2}{4!} {\phi}_1 *'
{\phi}_1 *' {\phi}_1 *' {\phi}_1\bigg],\nonumber
\end{eqnarray} 
where ${\phi}_1{\equiv}{\phi}_1(x^{NC})=\int
\frac{d^4p}{(2{\pi})^4}{\phi}_{1}(\vec{p})e^{-ipx^{NC}}={\phi}_1^{\dagger}$
and $*'$ is the canonical (or Moyal-Weyl) star product
\begin{equation} 
f *' g(x^{NC}) = e^{\frac{i}{2}{\theta}^{'2} B^{{\mu}{\nu}}
{\partial}_{\mu}^y {\partial}_{\nu}^z} f(y)g(z)|_{y=z=x^{NC}}
\label{star}
\end{equation} 
This is consistent with the commutation relation (\ref{comm1}) and
provides a nice check that that the canonical star product on the
sphere derived in \cite{dolan} (also given here by equation
(\ref{early})) reduces in the limit (\ref{contrast}) to the above
Moyal-Weyl product (\ref{star}). In the above, $B$ is the
antisymmetric tensor which can always be rotated such that the non
vanishing components are given by $B^{12}=-B_{21}=-1$ and
$B^{34}=-B_{43}=-1$.
 
In fact one can read immediately from the above effective action that
the planar contribution is quadratically divergent as it should be,
i.e.
\begin{eqnarray}
{\Delta}M^P=\frac{1}{64{\pi}^2}{\delta}M^P =
\int_{\sqrt{l}{\Lambda}^{'}{\rightarrow}{\infty}}
\frac{d^4k}{(2{\pi})^4}\frac{1}{k^2+M^2} =
\frac{1}{16{\pi}^2}l{\Lambda}^{'2}{\rightarrow}{\infty}, 
\end{eqnarray}
whereas the non-planar contribution is clearly finite 
\begin{eqnarray} 
{\Delta}M^{NP}(p)=\frac{1}{32{\pi}^2}{\delta}M^{NP}(p) &=&
\int_{\sqrt{l}{\Lambda}^{'}{\rightarrow}{\infty}}
\frac{d^4k}{(2{\pi})^4}\frac{e^{i{\theta}^{'2}\vec{p}B\vec{k}}}{k^2+M^2}
\nonumber\\
&=&\frac{1}{8{\pi}^2}\bigg[\frac{2}{E^2{{\theta'}^4}} + M^2
\ln({\theta'}^2 EM)\bigg], \quad {\rm where} \quad
E^{\nu}=B^{{\mu}{\nu}}P_{\mu}.
\end{eqnarray} 
This is the answer of \cite{mirase}: it is singular at $P=0$ as well
as at $\theta'=0$.

\subsection{A New Planar Limit With Strong Noncommutativity}
As explained earlier, the limit (\ref{flattening2}) possesses the
attractive feature that a momentum space cut-off is naturally built
into it. In addition to obtaining a noncommutative plane in the strict
limit, UV-IR mixing is completely absent. But while the new scaling is
simply stated, obtaining the corresponding field theory is somewhat
subtle. We will need to modify the Laplacian on the fuzzy sphere to
project our modes with momentum greater than $2\sqrt{l}$. In other
words, the noncommutative theory on a plane with UV cut-off $\theta$
is obtained not by flattening the full theory on the fuzzy sphere, but
only a ``low energy'' sector, corresponding to momenta upto $2
\sqrt{l}$.

In order to clarify the chain of arguments, we will first implement
naively the limit (\ref{flattening2}) and show that it corresponds to
a strongly noncommuting plane . Finite noncommuting plane is only
obtainable if we pick a specific low energy sector of the fuzzy sphere
before taking the limit as we will explain in the next section.

Our rule for matching the spectrum on the fuzzy sphere with that on
the noncommutative plane is the same as before, namely
$a(a+1)=R^2p_{a}^2$. However because of (\ref{flattening2}), the range
of $p_a^2$ is now from $0$ to $\frac{2l(2l+1)}{R^2} =
\frac{4}{{\theta}^2}$. The kinetic part of the action will scale in
the same way as in (\ref{dimensionanalysis0}), only now the momenta
$\vec{p}$'s in (\ref{dimensionanalysis0}) are restricted such that
$p{\leq}{\Lambda}$. With this scaling information, we can see that the
planar contribution to the $2$-point function is given by
\begin{equation} 
{\delta}m^P{\equiv}\frac{{\delta}{\mu}_l^{P}}{R^2} = \frac{4}{{\pi}^2} \int_{k{\leq}{\Lambda}}
\frac{d^4k}{k^2 + {\mu}^2_l}, \quad {\Lambda} = \frac{2}{\theta}.
\label{planar20}
\end{equation} 
We can similarly compute the non-planar contribution to the $2$-point
function using (\ref{appr1}). The motivation for using this
approximation is more involved and can be explained as follows. In the
planar limit $l,R{\rightarrow}{\infty}$, it is obvious that the
relevant quantum numbers $k_1$ and $p_1$ are in fact much larger
compared to $1$, i.e. $k_1{\sim}R p_{k_1}>>1$ and $p_1{\sim}R
p_{p_1}>>1$, since $R{\simeq}{\theta}l$. However (\ref{appr1}) can be
used only if $k_1,p_1<<l$, or equivalently $\frac{k_1}{l} =
\frac{2p_{k_1}}{\Lambda}<<1$ and $\frac{p_1}{l} = \frac{2
p_{p_1}}{\Lambda}<<1$. This is clearly true for small external momenta
$p_{k_1}$ and $p_{p_1}$, which is exactly the regime of interest in
order to see if there is UV-IR mixing. The condition for the
reliability of the approximation (\ref{appr1}) is then ${\theta}
p_{external}<<1$. We will sometimes refer to this condition as
``$\theta$ small'', the precise meaning of this phrase being
``momentum scale of interest is much smaller than $1/\theta$''. We
thus obtain
\begin{equation} 
{\delta}m^{NP}(k_1,p_1){\equiv}\frac{{\delta}{\mu}_l^{NP}}{R^2} =
8\int_{0}^{\Lambda} \int_{0}^{\Lambda} 
\frac{p_ap_bdp_adp_b}{p_a^2 + p_b^2 + {\mu}^2_l} P_{k_1}(1 -
\frac{{\theta}^2 p_a^2}{2}) P_{p_1}(1 - \frac{{\theta}^2
p_b^2}{2}).
\end{equation} 
Now the angles ${\nu}_a$'s of (\ref{formula}) are defined by
$\cos{\nu}_{a}=1-\frac{{\theta}^2p_a^2}{2}$, and since ${\theta}p<<1$,
these angles are still small. They are therefore given to the leading
order in $\theta p$ by ${\nu}_a={\theta}p_a+ \cdots$ where the
ellipsis indicate terms third order and higher in $\theta p$. By using
(\ref{formula}) we again have
\begin{equation} 
P_{Rp_{k_1}}(1 - \frac{{\theta}^2 p_a^2}{2}) = J_0(R {\theta}
p_{k_1}p_a) = \frac{1}{2{\pi}} \int_{0}^{2{\pi}} d{\phi}_a
e^{i{R}{\theta} \cos{\phi}_a p_{k_1} p_a}.
\label{well}
\end{equation} 
Using rotational invariance we can rewrite this as
\begin{equation} 
{\delta}m^{NP}(p) = \frac{2}{{\pi}^2} \int_{k{\leq}{\Lambda}}
\frac{d^4k}{k^2 + {\mu}^2_l} e^{iR{\theta}pBk}.
\label{nonplanar20}
\end{equation} 
One immediate central remark is in order: the noncommutative phase
contains now a factor $R{\theta}$ instead of the naively expected
factor of ${\theta}^2$. This is in contrast with the previous case of
canonical planar limit, where the strength of the noncommutativity
${\theta'}^2$ defined by the commutation relation (\ref{comm1}) is
exactly what appears in the noncommutative phase of
(\ref{nonplanar10}). In other words this naive implementation of
(\ref{flattening2}) yields in fact the strongly noncommuting plane
(\ref{or}) instead of (\ref{commutationrelation}). Also we can
similarly to the previous case put together the tree level action
(\ref{dimensionanalysis0}) with the one-loop contributions
(\ref{planar20}) and (\ref{nonplanar20}) to obtain the effective
action
\begin{eqnarray} 
&&\int_{{\Lambda}} \frac{d^4\vec{p}}{(2{\pi})^4} \frac{1}{2}
\bigg[\vec{p}^2 + {\mu}^2_l + \frac{g_4^2}{6} \big[2 \int_{{\Lambda}}
\frac{d^4k}{(2{\pi})^4} \frac{1}{k^2 + {\mu}^2_l} + \int_{{\Lambda}}
\frac{d^4k}{(2{\pi})^4} \frac{1}{k^2 + {\mu}^2_l} e^{iR{\theta} \vec{p} B
\vec{k}} \big] \bigg]|{\phi}_{3}(\vec{p})|^2 .
\label{effective}
\end{eqnarray} 
As before $g_4^2=8{\pi}^2{\lambda}_4$, whereas
${\phi}_3(\vec{p})=l^{3/2}{\phi}_2(\sqrt{l}\vec{p})$,
${\phi}_{2}(\vec{p}) = {4{\pi}}\sqrt{\frac{2}{l^3}}
{\phi}_{NC}(\frac{\vec{p}}{\sqrt{l}})$ with ${\phi}_{NC}(\vec{p})
{\equiv} {\phi}_{NC}^{p_ap_b{\phi}_a{\phi}_b} = R^4{\phi}^{abm_am_b}$
(in the metric ${\mathbb R}^2 \times {\mathbb R}^2$). It is not
difficult to see that the one-loop contributions ${\delta}m^P$ and
${\delta}m^{NP}(p)$ given in (\ref{planar20}) and (\ref{nonplanar20})
can also be given by the equations
\begin{eqnarray}
\bar{\Delta}m^{P} &=& \frac{l}{64{\pi}^2}{\delta}m^P =
\int_{\sqrt{l}{\Lambda}{\rightarrow}{\infty}}\frac{d^4k}{(2{\pi})^4}
\frac{1}{k^2 + l{\mu}^2_l}\nonumber\\ 
\bar{\Delta}m^{NP}(p) &=&
\frac{l}{32{\pi}^2}{\delta}m^{NP}(\frac{p}{\sqrt{l}}) =
\int_{\sqrt{l}{\Lambda}{\rightarrow}{\infty}} \frac{d^4k}{(2{\pi})^4}
\frac{1}{k^2 + l{\mu}^2_l} e^{i{\theta}^2 \vec{p} B \vec{k}}.
\end{eqnarray}
We have already computed that the leading terms in $\bar{\Delta}m^{P}$
and $\bar{\Delta}m^{NP}(p)$ are given by
\begin{eqnarray}
\bar{\Delta}m^{P} = \frac{l}{16{\pi}^2}\bigg[{\Lambda}^2 - {\mu}^2_l
  \ln(1+\frac{{\Lambda}^2}{{\mu}^2_l}\bigg],\;\;\bar{\Delta}m^{NP}(p) 
=\frac{1}{8{\pi}^2}\bigg[\frac{2}{E^2{{\theta}^4}} + l{\mu}^2_l 
\ln({\theta}^2 \sqrt{l}E{\mu}_l)\bigg], \quad {\rm where} \quad
E^{\nu}=B^{{\mu}{\nu}}p_{\mu}.\nonumber
\end{eqnarray}
Obviously then we obtain
\begin{equation} 
{\delta}m^P = 4\bigg[{\Lambda}^2 -
  {\mu}^2_lln(1+\frac{{\Lambda}^2}{{\mu}^2_l}\bigg], \quad
  {\delta}m^{NP}(p) = 4{\mu}^2_l \ln(l{\theta}^2E{\mu}_l).
\end{equation} 
If we now require the mass ${\mu}_l$ in (\ref{dimensionanalysis0}) to
scale as ${\mu}_l^2 = \frac{m^2}{l}$ (the reason will be clear
shortly), then one can deduce immediately that the planar contribution
${\delta}m^P$ is exactly finite equal to $4{\Lambda}^2$, whereas the
non-planar contribution ${\delta}m^{NP}(p)$ vanishes in the limit
$l{\rightarrow}{\infty}$.

Remark finally that despite the presence of the cut-off ${\Lambda}$ in
the effective action (\ref{effective}), this effective action can
still be obtained from quantizing
\begin{equation} 
\int d^4x \bigg[\frac{1}{2} ({\partial}_{\mu} {\phi}_3)^2 +
\frac{1}{2}{\mu}_l^2{\phi}_3^2 + \frac{g_4^2}{4!} {\phi}_3* {\phi}_3*
{\phi}_3* {\phi}_3\bigg],
\label{ncr4action}
\end{equation} 
only we have to regularize all integrals in the quantum theory with a
cut-off $\Lambda=2/\theta$. [${\phi}_3{\equiv}{\phi}_3(x^{F}) = \int
\frac{d^4p}{(2{\pi})^4} {\phi}_{3} (\vec{p}) e^{-ipx^{F}} =
{\phi}_3^{\dagger}$, and the star product $*$ is the Moyal-Weyl
product given in (\ref{star}) with the obvious substitution
${\theta}'{\rightarrow}R{\theta}$].

\subsection{A New Planar Limit With Finite Noncommutativity}

Neverthless, the action (\ref{effective}) can also be understood in
some way as the effective action on the noncommutative plane
(\ref{commutationrelation}) with finite noncommutativity equal to
${\theta}^2$. Indeed by performing the rescaling
$\vec{p}{\rightarrow}\frac{\vec{p}}{\sqrt{l}}$ we get
\begin{eqnarray} 
&&\int_{\sqrt{l}{\Lambda}} \frac{d^4\vec{p}}{(2{\pi})^4} \frac{1}{2}
\bigg[\vec{p}^2 + m^2 + \frac{g_4^2}{6} \big[2 \int_{\sqrt{l}{\Lambda}}
\frac{d^4k}{(2{\pi})^4} \frac{1}{k^2 + m^2} + \int_{\sqrt{l}{\Lambda}}
\frac{d^4k}{(2{\pi})^4} \frac{1}{k^2 + m^2} e^{i{\theta}^2 \vec{p} B
\vec{k}} \big] \bigg]|{\phi}_{2}(\vec{p})|^2 .
\label{effective1}
\end{eqnarray} 
We have already the correct noncommutativity $\theta^2$ in the phase
and the only thing which needs a new reintepretation is the fact that
the cut-off is actually given by
$\sqrt{l}{\Lambda}{\rightarrow}{\infty}$ and not by the finite cut-off
$\Lambda$. [Remark that if we do not reduce the cut-off
$\sqrt{l}{\Lambda}$ again to the finite value ${\Lambda}$, the physics
of (\ref{effective1}) is then essentially that of canonical
noncommutativity, i.e the limit (\ref{flattening2}) together with the
above rescaling of momenta is equivalent to the limit
(\ref{contrast})].

Now having isolated the $l$-dependence in the range of momentum space
integrals in the effective action (\ref{effective1}), we can argue
that it is not possible to get rid of this $l$-dependence merely by
changing variables. Actually, to correctly reproduce the theory on the
noncommutative ${\mathbb R}^4$ given by (\ref{flattening2}) and
(\ref{commutationrelation}), we will now show that one must start with
a modified Laplacian (or alternately propagator) on the fuzzy space
\cite{denj}. For this, we replace the Laplacian ${\Delta} =
[L_i^{(a)},[L_i^{(a)},..]]$ (see equation (\ref{lapl}), $a=1,2$) on each fuzzy
sphere which has the canonical obvious spectrum $k(k+1)$,
$k=0,...,2l$, with the modified Laplacian
\begin{eqnarray} 
{\Delta}_j = {\Delta} + \frac{1}{\epsilon}(1 - P_j).
\label{presc}
\end{eqnarray} 
Here $P_j$ is the projector on all the modes associated with the
eigenvalues $k=0,...,j$, i.e.
\begin{eqnarray} 
P_j = \sum_{k=0}^{j}\sum_{m=-k}^k|k,m \rangle \langle k,m|, \nonumber
\end{eqnarray} 
The integer $j$ thus acts as an intermediate scale, and using the
modified propagator gives us a low energy sector of the full
theroy. We will fix the integer $j$ shortly.

With this modified Laplacian, modes with momenta larger than $j$ do
not propagate: as a result, they make no contribution in momentum sums
that appear in internal loops. In other words, summations like
$\sum_{0}^{2l}$ (which go over to integrals with range
$\int_0^\Lambda$) now collapse to $\sum_{0}^{j}$ (where the integrals
now are of the range $\int_0^{{\Lambda}_j}$, with ${\Lambda}_j =
\frac{j}{2l} \Lambda$).

The new flattening limit is now defined as follows: start with the
theory on $S_F^2 \times S_F^2$, but with the modified propagator
(\ref{presc}). First take $\epsilon \rightarrow 0$, then $R,l
\rightarrow \infty$ with $\theta = R/l$ fixed. This gives us the
effective action (\ref{effective1}) but with with momentum space
cut-off $\sqrt{l}{\Lambda}_j=\frac{j}{2\sqrt{l}} \Lambda$ ,i.e
\begin{eqnarray} 
&&\int_{\sqrt{l}{\Lambda}_j} \frac{d^4\vec{p}}{(2{\pi})^4} \frac{1}{2}
\bigg[\vec{p}^2 + m^2 + \frac{g_4^2}{6} \big[2 \int_{\sqrt{l}{\Lambda}_j}
\frac{d^4k}{(2{\pi})^4} \frac{1}{k^2 + m^2} + \int_{\sqrt{l}{\Lambda}_j}
\frac{d^4k}{(2{\pi})^4} \frac{1}{k^2 + m^2} e^{i{\theta}^2 \vec{p} B
\vec{k}} \big] \bigg]|{\phi}_{2}(\vec{p})|^2 
\label{effective2}
\end{eqnarray} 
This also tells us that the correct choice of the intermediate scale
is $j =[2 \sqrt{l}]$ for which $\sqrt{l}{\Lambda}_j={\Lambda}$. For
this value of the intermediate cut-off, we obtain the noncommutative
${\mathbb R}^4$ given by (\ref{flattening2}) and
(\ref{commutationrelation}) .

By looking at the product of two functions of the fuzzy sphere, we can
understand better the role of the intermediate scale $j
(=[2\sqrt{l}])$. The fuzzy spherical harmonics $T_{l_a m_a}$ go over
to the usual spherical harmonics $Y_{l_a m_a}$ in the limit of large
$l$, and so does their product, provided their momenta are
fixed. Alternately, the product of two fuzzy spherical harmonics $T$'s
is ``almost commutative'' (i.e. almost the same as that of the
corresponding $Y$'s) if their angular momentum is small compared to
the maximum angular momentum $l$, whereas it is ``strongly
noncommutative'' (i.e. far from the commutative regime) if their
angular momenta are sufficiently large and comparable to $l$. The
intermediate cut-off tells us precisely where the product goes from
one situation to the other: Working with fields having momenta much
less than $[2\sqrt{l}]$ leaves us in the approximately commutative
regime, while fields with momenta much larger than $[2\sqrt{l}]$ take
us in the strongly noncommutative regime. In other words, the
intermediate cut-off tells us where commutativity and noncommutativity
are in delicate balance. Indeed by writing (\ref{effective2}) in the
form

\begin{eqnarray} 
&&\int_{\sqrt{l}{\Lambda}_j} \frac{d^4\vec{p}}{(2{\pi})^4} \frac{1}{2}
\bigg[\vec{p}^2 + m^2 + \frac{g_4^2}{6} \big[2 \int_{\sqrt{l}{\Lambda}_j}
\frac{d^4k}{(2{\pi})^4} \frac{1}{k^2 + m^2} + \int_{\sqrt{l}{\Lambda}_j}
\frac{d^4k}{(2{\pi})^4} \frac{1}{k^2 + m^2} e^{i{\theta}^2 \vec{p} B
\vec{k}} \big] \bigg]|{\phi}_{2}(\vec{p})|^2{\equiv}\nonumber\\
&&\int_{\Lambda} \frac{d^4\vec{p}}{(2{\pi})^4} \frac{1}{2}
\bigg[\vec{p}^2 + {\mu}^2_{l,j}+ \frac{g_4^2}{6} \big[2 \int_{{\Lambda}}
\frac{d^4k}{(2{\pi})^4} \frac{1}{k^2 + {\mu}^2_{l,j}} + \int_{{\Lambda}}
\frac{d^4k}{(2{\pi})^4} \frac{1}{k^2 + {\mu}^2_{l,j}} e^{i(\frac{j}{2\sqrt{l}})^2{\theta}^2 \vec{p} B
\vec{k}} \big] \bigg]|{\phi}^{(j)}_{3}(\vec{p})|^2.\nonumber
\label{effective3}
\end{eqnarray}
[${\mu}^2_{l,j}=l{\mu}^2_l(\frac{2\sqrt{l}}{j})^2$ ,
${\phi}^{(j)}_3(\vec{p})=(\frac{j}{2\sqrt{l}})^3{\phi}_2(\frac{j}{2\sqrt{l}}\vec{p})$
, ${\phi}^{(2l)}_3{\equiv}{\phi}_3$]. For $j<<[2\sqrt{l}]$,
$(\frac{j}{2\sqrt{l}})^2{\theta}^2{\rightarrow}0$ and this is the
effective action on a commutative ${\mathbb R}^4$ with cut-off
${\Lambda}=2/\theta$. For $j>>[2\sqrt{l}]$ this effective action
corresponds to canonical noncommutativity if we insist on the first
line above as our effective action or to strongly noncommuting
${\mathbb R}^4$ if we consider instead the effective action to be
given by the second line. For the value $j=[2\sqrt{l}]$, where we
obtain the noncommutative ${\mathbb R}^4$ given by (\ref{flattening2})
and (\ref{commutationrelation}), there seems to be a balance between
the above two situations and one can also expect the UV-IR mixing to
be smoothen out.

To show this we write first the one-loop planar and non-planar
contributions for $j=[2\sqrt{l}]$ , viz
\begin{eqnarray}
{\Delta}m^P&=&\int_{{\Lambda}}
\frac{d^4k}{(2{\pi})^4} \frac{1}{k^2 + m^2}~,~
{\Delta}m^{NP}(p)=\int_{{\Lambda}}
\frac{d^4k}{(2{\pi})^4} \frac{1}{k^2 + m^2} e^{i{\theta}^2 \vec{p} B
\vec{k}}. \nonumber
\end{eqnarray}
We can evaluate these integrals by introducing a Schwinger parameter
$(k^2+m^2)^{-1}=\int
d{\alpha}exp\bigg(-\alpha(k^2+m^2)\bigg)$. Explicitly, we obtain for
the planar contribution
\begin{eqnarray}
{\Delta}m^P &=& \frac{1}{16{\pi}^2}\bigg[ -{\Lambda}^2\int
  \frac{d\alpha}{\alpha}e^{-\alpha(m^2+{\Lambda}^2)} + \int
  \frac{d\alpha}{{\alpha}^2}e^{ -\alpha m^2}\bigg(1 - e^{-\alpha
  {\Lambda}^2}\bigg)\bigg] \nonumber\\
&=&\frac{1}{16{\pi}^2}\bigg[{\Lambda}^2+m^2 \ln\frac{m^2}{m^2+{\Lambda}^2}\bigg].
\end{eqnarray}

Obviously the above planar function diverges quadratically as
${\Lambda}^2$ when $\theta{\rightarrow}0$, i.e. the noncommutativity
acts effectively as a cut-off.

Next we compute the non-planar integral. To this end we introduce as
above a Schwinger parameter and rewrite the integral as follows

\begin{eqnarray}
{\Delta}m^{NP}(p)&=&
\frac{1}{16{\pi}^4}\int_{0}^{\infty}
d{\alpha}e^{-{\alpha}m^2-\frac{{\theta}^4E^2}
  {4{\alpha}}}\int_{{\Lambda}}d^4k e^{-{\alpha}\big[\vec{k} -
    \frac{i{\theta}^2}{2{\alpha}}\vec{E}\big]^2}\nonumber\\ 
&=&\frac{1}{16{\pi}^4} \sum_{r=0}^{\infty}({\theta}^2)^r
\sum_{s=0}^{[\frac{r}{2}]}\frac{i^{r-s}}{s!(r-2s)!}\bigg[\int_{0}^{\infty} 
d{\alpha}(\frac{E^2}{4i\alpha})^se^{-{\alpha}m^2-\frac{{\theta}^4E^2}
{4{\alpha}}}\bigg[\int_{{\Lambda}}d^4k e^{-\alpha
k^2}(\vec{k}\vec{E})^{r-2s}\bigg]\bigg]~,~E^{\nu}=B^{\mu
\nu}p^{\mu}.\nonumber
\end{eqnarray}
In above we have also used the fact that $\theta$ is small in the
sense we explained earlier (i.e. $E\theta<<1$) and in accordance with
\cite{wulkenhaar} to expand the second exponential around
$\theta=0$. This is also because the cut-off ${\Lambda}$ is inversely
proportional to $\theta$. [In the last line we used the identity
$\sum_{p=0}^{\infty}\sum_{q=0}^pA_{q,p-q}=\sum_{r=0}^{\infty}\sum_{s=0}^{[\frac{r}{2}]}A_{s,r-2s}$,
$[\frac{r}{2}]=\frac{r}{2}$ for $r$ even and
$[\frac{r}{2}]=\frac{r-1}{2}$ for $r$ odd] . It is not difficult to
argue that the inner integral above vanishes unless $r$ is even. Using
also the fact that the cut-off ${\Lambda}$ is rotationally invariant
one can evaluate the inner integral as follows. We have
\begin{eqnarray}
\int_{{\Lambda}}d^4ke^{-\alpha k^2}(\vec{k}\vec{E})^{n}
&=& 4{\pi}^2E^n(n-1)!!\bigg[ \frac{1}{(2\alpha)^{\frac{n}{2} + 2}} -
  {\Lambda}^ne^{-\alpha{\Lambda}^2}
  \sum_{q=-1}^{\frac{n}{2}}\frac{1}{(n-2q)!!}
  \frac{1}{(2\alpha)^{q+2}}\frac{1}{{\Lambda}^{2q}}\bigg],\nonumber 
\end{eqnarray}
where $n$ is an even number given by $n=r-2s$. 

We can now put the above non-planar function in the form
\begin{eqnarray}
&&{\Delta}m^{NP}(p)
=\frac{1}{16{\pi}^2}\sum_{N=0}^{\infty}\frac{1}{N!}\big(\frac{{\theta}^2E}{2}\big)^{2N}\int
\frac{d\alpha}{{\alpha}^{N+2}}e^{-\alpha
m^2-\frac{{\theta}^4E^2}{4\alpha}}\sum_{M=0}^{N}C_N^M(-1)^M\bigg[1-\sum_{P=0}^{M+1}\frac{(\alpha
{\Lambda}^2)^P}{P!}e^{-\alpha
{\Lambda}^2}\bigg].\label{uvir}\nonumber\\
\end{eqnarray}
[$C_N^M=\frac{N!}{M!(N-M)!}$]. The first term in this expansion
corresponds exactly to the case of canonical noncommutativity where
instead of ${\Lambda}$ we have no cut-off, i.e.
\begin{eqnarray}
{\Delta}m^{NP}(p)
&=&\frac{1}{16{\pi}^2}\sum_{N=0}^{\infty}\frac{1}{N!}\big(\frac{{\theta}^2E}{2}\big)^{2N}\int
\frac{d\alpha}{{\alpha}^{N+2}}e^{-\alpha
m^2-\frac{{\theta}^4E^2}{4\alpha}}\sum_{M=0}^{N}C_N^M(-1)^M+...\nonumber\\
&=& \frac{1}{8{\pi}^2}\bigg[\frac{2}{{\theta}^4E^2} + m^2
  \ln(m{\theta}^2E)\bigg] +...{\equiv} \frac{1}{16{\pi}^2}I^{(2)}(m^2,
\frac{{\theta}^4E^2}{4})+...\nonumber
\end{eqnarray}
As expected this term provides essentially the canonical UV-IR
mixing. As it turns out this singular behaviour is completely
regularized by the remaining $N=0$ term in (\ref{uvir}), i.e.
\begin{eqnarray}
{\Delta}m^{NP}(p)&=&\frac{1}{16{\pi}^2}I^{(2)}(m^2,\frac{{\theta}^4E^2}{4})
+\frac{1}{16{\pi}^2}\int \frac{d\alpha}{{\alpha}^{2}}e^{-\alpha
m^2-\frac{{\theta}^4E^2}{4\alpha}}\bigg[-\sum_{P=0}^{1}\frac{(\alpha
{\Lambda}^2)^P}{P!}e^{-\alpha {\Lambda}^2}\bigg]+......\nonumber\\
&=& \frac{1}{16{\pi}^2} I^{(2)}(m^2,\frac{{\theta}^4E^2}{4}) -
\frac{1}{16{\pi}^2} \bigg[I^{(2)}(m^2 +
  {\Lambda}^2,\frac{{\theta}^4E^2}{4}) +
  {\Lambda}^2I^{(1)}(m^2+{\Lambda}^2, \frac{{\theta}^4E^2}{4})\bigg] +
...\label{equation}
\end{eqnarray}
The integrals $I^{(L)}(x,y)$ are given essentially by Hankel functions , viz
\begin{eqnarray}
&&I^{(1)}(x,y) = \int_{0}^{\infty}
  \frac{d{\alpha}}{{\alpha}}e^{-x\alpha-\frac{y}{\alpha}} =
  \frac{1}{2}\bigg[i{\pi}H_0^{(1)}(2i\sqrt{xy})
  +h.c. \bigg]\nonumber\\ 
&&~I^{(L)}(x,y) = \int_{0}^{\infty}
  \frac{d{\alpha}}{{\alpha}^L}e^{-x\alpha-\frac{y}{\alpha}} =
  \frac{1}{2}\bigg[\frac{i{\pi}}{L-1}(\frac{x}{y})^{\frac{L-1}{2}}
  \sqrt{xy}e^{\frac{iL{\pi}}{2}} \big[H_{L-2}^{(1)}(2i\sqrt{xy}) +
  H_L^{(1)}(2i\sqrt{xy})\big] + h.c. \bigg]~,~L>1.\nonumber
\end{eqnarray}
Hankel functions admit the series expansion $H_0^{(1)}(z)
=\frac{2i}{\pi} \ln z+...$ and $H_{\nu}^{(1)}(z) =
-\frac{i(\nu-1)!}{\pi}(\frac{2}{z})^{\nu}+..$ for $\nu>0$ when
$z{\longrightarrow}0$. In this case the mass $m$ and the external
momentum $E$ are both small compared to the cut-off $\Lambda=2/\theta$
and thus the dimensionless parameters $z{\equiv}\sqrt{xy} =
2\frac{m}{\Lambda}\frac{E}{\Lambda}$ or $z{\equiv}\sqrt{xy} = 2\sqrt{1
+ \frac{m^2}{{\Lambda}^2}}\frac{E}{\Lambda}$ are also small, in other
words we can calculate for example $I^{(1)}(x,y) = -2
\ln(2\sqrt{xy})$, $I^{(2)}(x,y) = 2x \ln(2\sqrt{xy}) + \frac{1}{y}$
and $I^{(L)}(x,y) = \frac{(L-2)!}{y^{L-1}}[1-\frac{xy}{(L-2)(L-1)}]$
for $L{\geq}3$. Thus the first term $N=0$ in the above sum ( i.e
euqation (\ref{equation})) is simply given by
\begin{eqnarray}
{\Delta}m^{NP}(p)=-\frac{m^2}{16{\pi}^2}ln(1+\frac{{\Lambda}^2}{m^2})+....
\end{eqnarray}
As one can see it does not depend on the external momentum $p$ at
all. In the commutative limit $\theta{\rightarrow}0$, this
diverges logarithmically as $ln{\Lambda}$ which is subleading compared
to the quadratic divergence of the planar function. Higher corrections
can also be computed and one finds essentially an expansion in
$\frac{{\Lambda}{\theta}^2E}{2}=E\theta=2\frac{E}{\Lambda}$ given by
\begin{eqnarray}
{\Delta}m^{NP}(p) &=& -\frac{m^2}{16{\pi}^2}
\ln(1+\frac{{\Lambda}^2}{m^2}) \nonumber\\
&+& \frac{{\Lambda}^2}{16{\pi}^2}I^{(1)}(x,y)
\sum_{p=2}^{\infty}\frac{1}{p!} \bigg(\frac{{\Lambda}{\theta}^2E}{2}
\bigg)^{2(p-1)}{\eta}_{p-1,p-2} + \frac{1}{16{\pi}^2}I^{(2)}(x,y)
\sum_{p=2}^{\infty}\frac{1}{p!} \bigg(\frac{{\Lambda}{\theta}^2E}{2}
\bigg)^{2p}{\eta}_{p,p-2} \nonumber\\
&+& \frac{1}{16{\pi}^2} \sum_{N=1}^{\infty}
\bigg(\frac{{\theta}^4E^2}{4} \bigg)^NI^{(N+2)}(x,y)
\sum_{p=2}^{\infty}\frac{1}{p!}
\bigg(\frac{{\Lambda}{\theta}^2E}{2}\bigg)^{2p}{\eta}_{p+N,p-2}.\nonumber 
\end{eqnarray}
[$x=m^2+{\Lambda}^2$, $y=\frac{{\theta}^4E^2}{4}$, ${\eta}_{p+N,p-2} =
  \sum_{M=0}^{p-2}\frac{(-1)^{M}}{M!(p+N-M)!}$]. It is not difficult
  to find that the leading terms in the limit of small external
  momenta (i.e. $E/\Lambda<<1$) are effectively given by
\begin{eqnarray}
{\Delta}m^{NP}(p) &=& -\frac{m^2}{16{\pi}^2} \ln\bigg(1 +
\frac{{\Lambda}^2}{m^2}\bigg) - \frac{E^2}{4{\pi}^2}
\ln\bigg(4\frac{E}{\Lambda}\sqrt{1 + \frac{m^2}{{\Lambda}^2}}\bigg)
\bigg[1 + O\bigg(\frac{E^2}{{\Lambda}^2}\bigg)\bigg] +
\frac{E^2}{8{\pi}^2} \bigg[1 +
  O^{'}\bigg(\frac{E^2}{{\Lambda}^2}\bigg)\bigg]. \label{np1}\nonumber\\
\end{eqnarray}
Clearly in the strict limit of small external momenta when
$E{\rightarrow}0$, we have $E^2 \ln E{\rightarrow}0$ and the
non-planar contribution does not diverge (only the first term in
(\ref{np1}) survives this limit as it is independent of $E$) and hence
there is no UV-IR mixing. The limit of zero noncommutativity is
singular but now this divergence has the nice interpretation of being
the divergence recovered in the non-planar $2-$point function when the
cut-off ${\Lambda} = \frac{2}{\theta}$ is removed. This divergence is
however logarithmic and therefore is sub-leading compared to the
quadratic divergence in the planar part.

The effective action (\ref{effective2}) with $j=[2\sqrt{l}]$ can be
obviously obtained from quantizing the action (\ref{ncr4action}) with
the replacements ${\mu}_l^2{\rightarrow}m^2$,
${\phi}_3{\rightarrow}{\phi}_2{\equiv}{\phi}_2(X^{NC}) = \int
\frac{d^4p}{(2{\pi})^4} {\phi}_{2} (\vec{p}) e^{-ipX^{NC}} =
{\phi}_2^{\dagger}$ and where as before we have to regularize all
integrals in the quantum theory with a cut-off $\Lambda=2/\theta$. The
star product $*$ is the Moyal-Weyl product given in (\ref{star}) with
the substitutions ${\theta}'{\rightarrow}{\theta}$,
$x^{NC}{\longrightarrow}X^{NC}$. This effective action can also be
rewritten in the form
\begin{equation} 
\int d^4x \bigg[\frac{1}{2} {\partial}_{\mu}{\phi}_2*_{\Lambda}
  {\partial}_{\mu}{\phi}_2 + \frac{1}{2} m^2{\phi}_2*_{\Lambda}
  {\phi}_2 + \frac{g_4^2}{4!}  {\phi}_2* {\phi}_2*
  {\phi}_2*{\phi}_2\bigg],
\label{action1}
\end{equation} 
which is motivated by the fact that the {\it effective} star product
defined by
\begin{eqnarray}
f*_{\Lambda}g(X^{NC}) &=& \int_{{\Lambda}} \frac{d^4p}{(2{\pi})^4}
f(\vec{p})\int_{{\Lambda}} \frac{d^4k}{(2{\pi})^4} g(\vec{k})
e^{-ipX^{NC}}*e^{-ikX^{NC}} \nonumber\\
&=& \int d^4y'd^4z' {\delta}^4_{\Lambda}(y') {\delta}^4_{\Lambda}(z')
f(y-y')*g(z-z')|_{y=z=X^{NC}},
\label{prod}
\end{eqnarray}
is such that $\int d^4xf*_{\Lambda} g(x) = \int_{{\Lambda}}
\frac{d^4p} {(2{\pi})^4} f(\vec{p}) g(-\vec{p})$. The distribution
${\delta}^4_{\Lambda}(y')$ is not the Dirac delta function
${\delta}^4(y')$ but rather ${\delta}^4_{\Lambda}(y') =
\int_{{\Lambda}} \frac{d^4p} {(2{\pi})^4} e^{-ipy'}$,
i.e. ${\delta}^4_{\Lambda}(y')$ tends to the ordinary delta function
in the limit ${\Lambda}{\rightarrow}{\infty}$ of the commutative plane
where the above product (\ref{prod}) also reduces to the ordinary
point-wise multiplication of functions. If the cut-off ${\Lambda}$ was
not correlated with the non-commutativity parameter ${\theta}$, then
the limit ${\Lambda}{\rightarrow}{\infty}$ would had corresponded to
the limit where the product (\ref{prod}) reduces to the Moyal-Weyl
product given in equation (\ref{star}). This way of writing the
effective action (i.e. (\ref{action1})) is to insist on the fact that
all integrals are regularized with a cut-off ${\Lambda}=2/\theta$. In
other words the above new star product which appears only in the
kinetic part of the action is completely equivalent to a sharp cut-off
${\Lambda}$ and yields therefore exactly the propagator (\ref{presc})
with which only modes ${\leq}{\Lambda}$ can propagate.

We should also remark here regarding non-locality of the star product
(\ref{prod}). At first sight it seems that this non-locality is more
severe in (\ref{prod}) than in (\ref{star}), but as it turns out this
is not entirely true: in fact the absence of the UV-IR mixing in this
product also suggests this. In order to see this more explicitly we
first rewrite (\ref{prod}) in the form
\begin{eqnarray}
&& f*_{{\Lambda}}g(X^{NC}) = \int d^4y' d^4z' f(y') g(z')
  K_{\Lambda}(y',z';X^{NC}) \nonumber\\
&& K_{\Lambda}(y',z';X^{NC}) = {\delta}^4_{\Lambda}(y-y')*
  {\delta}^4_{\Lambda}(z-z')|_{y=z=X^{NC}}. \nonumber
\end{eqnarray}
The kernel $K_{\Lambda}$ can be computed explicitly and is given
by
\begin{equation} 
K_{\Lambda}(y',z';X^{NC}) = \int_{{\Lambda}} \frac{d^4k}{(2{\pi})^4}
{\delta}^4_{\Lambda}(X^{NC}-y' + \frac{{\theta}^2}{2}Bk) e^{ik(z' -
  X^{NC})}. \nonumber
\end{equation} 
For the moment, let us say that $\Lambda$ and $\theta$ are
unrelated. Then, taking ${\Lambda}$ to infinity gives \cite{mirase,dn}
\begin{equation} 
K(y',z';X^{NC}) = \frac{16}{{\theta}^8 \det B} \frac{1}{(2{\pi})^4}
e^{\frac{2i}{{\theta}^2}(z' - X^{NC})B^{-1}(y' - X^{NC})}. \nonumber
\end{equation} 
If we have for example two functions $f$ and $g$ given by
$f(x)={\delta}^4(x-p)$ and $g(x)={\delta}^4(x-p)$, i.e. they are
non-zero only at one point $p$ in space-time, their star product which
is clearly given by the kernel $K(p,p;X^{NC})$ is non-zero everywhere
in space-time. The fact that $K$ is essentially a phase is the source
of the non-locality of (\ref{star}) which leads to the UV-IR mixing.

On the other hand the kernel $K_{\Lambda}(p,p;X^{NC})$ with finite
${\Lambda}$ can be found in two dimensions (say) to be given by
\begin{equation} 
K_{\Lambda}(p,p;X^{NC}) = \frac{1}{{\pi}^2{\theta}^4}
\int_{-{\theta}}^{{\theta}} da{\delta}_{\Lambda}(a +
L_1) e^{\frac{2i}{{\theta}^2}L_2a}
\int_{-{\theta}}^{{\theta}} db{\delta}_{\Lambda}(b +
L_2) e^{-\frac{2i}{{\theta}^2}L_1b},\nonumber
\label{effestar}
\end{equation} 
with $L_a=X_a^{NC}-p_a$, $a=1,2$. If we now make the approximation to
drop the remaining ${\Lambda}$ (since the effects of this cut-off were
already taken anyway) one can see that the above integral is non-zero
only for $-{\theta}+p_1{\leq}X_1^{NC}{\leq}{\theta}+p_1$ and
$-{\theta}+p_2{\leq}X_2^{NC}{\leq}{\theta}+p_2$ simultaneously. In
other words the star product $K_{\Lambda}(p,p;X^{NC})$ of $f(x)$ and
$g(x)$ is also localized around $p$ within an error ${\theta}$ and is
equal to $\frac{1}{{\pi}^2{\theta}^4}$ there . The star product
(\ref{prod}) is therefore effectively local.

Final remarks are in order. First we note that the effective star
product (\ref{prod}) leads to an effective commutation relations
(\ref{commutationrelation}) in which the parameter ${\theta}^2$ is
multiplied by an overall constant equal to $\int d^4y' d^4z'
{\delta}^4_{\Lambda}(y'){\delta}^4_{\Lambda}(z')$, we simply skip the
elementary proof. Remark also that this effective star product is
non-associative as one should expect since it is for all practical
purposes equivalent to a non-trivial sharp momentum cut-off ${\Lambda}$
\cite{ydri}.

The last remark is to note that the prescription (\ref{presc}) can
also be applied to the canonical limit of large stereographic
projection of the spheres onto planes, and in this case one can also
obtain a cut-off ${\Lambda}' = \frac{2}{{\theta}'}$ with $j$ fixed as
above such that $j = [2\sqrt{l}]$. The noncommutative plane
(\ref{comm1}) defined in this way is therefore completely equivalent
to the above noncommutative plane (\ref{commutationrelation}).

\subsection{The Continuum Planar Limit of the $4-$Point Function }

We now undertake the task of finding the continuum limit of the above
$4$-point function (equations (\ref{1000}) and (\ref{1002})) which we
expect to correspond to the $4$-point function on the
noncommutative ${\mathbb R}^4$. This expectation is motivated of course by the
result of the last sections on the $2$-point function. As it turns out this is also the case here and as an explicit example we work out the continuum flattening limit of the planar amplitudes  .

The planar diagrams are
${\delta}{\lambda}_4^{(1)}$ and ${\delta}{\lambda}_4^{(4)}$.  First,
let us recall that in above the indices $4$ and $6$ refer to internal
momenta whereas $1$, $2$, $3$ and $5$ refer to external momenta. Next,
since we are interested in the planar limits (in which
$R,l{\rightarrow}{\infty}$) of the $4$-point function, we can use the
asymptotic formula
\begin{equation} 
\left\{
\begin{array}{ccc}
a&b  & c\\
d+l&e+l&f+l
\end{array}
\right\} = \frac{(-1)^{a+b+d+e}}{\sqrt{(2l+1)(2c+1)}}
C^{cd-e}_{af-ebd-f}, \quad l{\rightarrow}{\infty},
\label{key}
\end{equation} 
which allows us to approximate in the limit the ``fuzzy delta'' function (\ref{delta})
as follows:
\begin{eqnarray}
{\delta}_k(1235) &=& (2l+1)(2k+1)(-1)^{k_1+k_2+k_3+k_5+m}
{\delta}_{m_1+m_2+m_3+m_5,0} \nonumber\\
&{\times}& \left\{
\begin{array}{ccc}
k_1&k_2  & k\\
m_2+l&-m_1+l&l
\end{array}
\right\} \left\{
\begin{array}{ccc}
k_3&k_5  & k\\
m_5+l&-m_3+l&l
\end{array}
\right\}.
\label{app}
\end{eqnarray}
We have also used the properties of the Clebsch-Gordan coefficients to
obtain the selection rule $m=m_1+m_2=-m_3-m_5$, thus justifying the
name. The next selection rule comes from the fact that the function
$E^{k_4k_6}_{k_1k_2}(k)E^{k_4k_6}_{k_3k_5}(k)$ in the planar diagrams
(\ref{planar4}) is proportional in the large $l$ limit (by virtue of
equation (\ref{key})) to $C_{k_10k_20}^{k0} C_{k_30k_50}^{k0}
(C_{k_40k_60}^{k0})^2$, whereas on the other hand these Clebsch-Gordan
coefficients are such that $C^{c0}_{a0b0}{\neq}0$ only if
$a+b+c$=even. This means in particular that $k + k_4 + k_6=~{\rm
even}$, $k + k_1 + k_2=~{\rm even}$ and $ k + k_3 + k_5=~{\rm even}$,
and hence one can argue in different ways that one can  have for example
\begin{equation} 
k_1 + k_2=k_3 + k_5, \quad k_1+k_2=k_4+k_6, \quad k_3+k_5=k_4+k_6.
\label{law}
\end{equation} 
For obvious reasons we will only focus on this sector. As a
consequence of these rules, the planar graphs ${\nu}_i^{(1)}$ and
${\nu}_i^{(4)}$ are equal. Indeed for large $l$, one can easily show
that these diagrams take the form
\begin{eqnarray}
{\nu}_i^{(1)} = {\nu}_i^{(4)} &{\simeq}& (2l+1)^3(-1)^{m_1+m_2}
{\delta}_{m_1+m_2+m_3+m_5,0} {\delta}_{m_1+m_2+m_4+m_6,0}
{\delta}_{k_1+k_2,k_3+k_5} {\delta}_{k_1+k_2,k_4+k_6}\nonumber\\
&{\times}& a_k(1235){\cal S}(46;1235) \quad {\rm where} \nonumber\\ 
{\cal S}(46;1235) &=& \sum_{k}(2k+1) \left\{
\begin{array}{ccc}
k_4&k_6  & k\\
l&l&l
\end{array}
\right\}^2 \left\{
\begin{array}{ccc}
k_1&k_2  & k\\
l&l&l
\end{array}
\right\}^2 \left\{
\begin{array}{ccc}
k_3&k_5  & k\\
l&l&l
\end{array}
\right\}^2,\
\label{4335}
\end{eqnarray}
where $a_k(1235) = \sqrt{\prod_{i=1}^{2,3,5}(2k_i+1)}$. As in the case
of the $2$-point function we have assumed that the external momenta
$k_1$, $k_2$, $k_3$ and $k_5$ are such that $k_i<<l$, $i=1,2,3,5$. It
is also expected that the approximation sign becomes an exact equality
only in the strict limit. Furthermore from the properties of the
$6j$-symbols, only the values $0{\leq}k{\leq}k_4+k_6$ will contribute
to the sum $\sum_{k}$. Lastly we have also invoked in (\ref{4335}) the
fact that for each fixed pair $(k_4,k_6)$ which is integrated over in
(\ref{1002}) the azimuth numbers $(m_4,m_6)$, although they are
already summed over, conspire such that their sum is
$m_4+m_6=-m_1-m_2$. From \cite{VMK} we can now use the identity
\begin{equation} 
\left\{\begin{array}{ccc}
        k_4&k_6  & k\\
          l&l&l
         \end{array} \right\}^2 = \sum_{X_1}(-1)^{X_1}(2X_1+1) \left\{
\begin{array}{ccc}
  k_4&l  & l\\
  X_1&l&l
\end{array}
\right\} \left\{
\begin{array}{ccc}
k_6&l  & l\\
X_1&l&l
\end{array}
\right\} \left\{
\begin{array}{ccc}
k&l  & l\\
X_1&l&l
\end{array}
\right\},
\label{x1}
\end{equation} 
etc. The delta function ${\delta}_{k_1+k_2,k_4+k_6}$ makes it safe to
treat the internal momenta $k_4$ and $k_6$ as if they were small
(recall that $k_4$ and $k_6$ are non-negative integers), and
$0{\leq}k{\leq}k_4+k_6$ means that $k$ can be treated as small as
well. One can therefore use the result (\ref{appr1}) to rewrite the
above equation as
\begin{eqnarray} 
\left\{\begin{array}{ccc}
 k_4&k_6  & k\\
 l&l&l
\end{array} \right\}^2 = \frac{1}{(2l+1)^3} \sum_{X_1=0}^{2l} (2X_1+1)
 P_{k_4}\left(1 - \frac{X_1^2}{2l^2} \right) P_{k_6}\left(1 -
 \frac{X_1^2}{2l^2}\right) P_{k}\left(1 -
 \frac{X_1^2}{2l^2}\right),\nonumber 
\end{eqnarray} 
etc. As we have already established, in the large $l$ limit we can
approximate this sum by the integral
\begin{equation} 
\left\{\begin{array}{ccc}
  A&B  & k\\
  l&l&l
\end{array} \right\}^2 = \frac{2R^2}{(2l+1)^3} \int_{0}^{{\Lambda}}
  p_{x_1}dp_{x_1} P_{A}\left(1 - \frac{{\theta}^2p^2_{x_1}}{2}\right)
  P_{B}\left(1 - \frac{{\theta}^2p^2_{x_1}}{2}\right) P_{k}\left(1 -
  \frac{{\theta}^2p^2_{x_1}}{2}\right). 
\end{equation} 
with $(A,B) = (k_1,k_2),(k_3,k_5)$ and $(k_4,k_6)$. We are obviously
using the flattening limit (\ref{flattening2}), i.e.  ${\theta} =
\frac{R}{\sqrt{l(l+1)}}$, ${\Lambda} = \frac{2}{{\theta}}$ for reasons
which will become self-evident shortly. Using the result (\ref{well})
we have
\begin{equation} 
P_{A}\left(1 - \frac{{\theta}^2p_{x_1}^2}{2}\right) P_{B}\left(1 -
\frac{{\theta}^2p^2_{x_1}}{2}\right) = \frac{1}{(2{\pi})^2} \int
d{\phi}_Ad{\phi}_B e^{iR{\theta}\vec{p}_{x_1}{\wedge}(\vec{p}_{A} +
\vec{p}_{B})}. 
\label{Int0}
\end{equation} 
where $\vec{p}_A{\wedge}\vec{p}_B =
B^{{\mu}{\nu}}p_A^{\mu}p_{B}^{\nu}$, with $B^{12}=-1$, and ${\phi}_A$,
${\phi}_B$ have the interpretation of angles between $\vec{p}_A$ and
$\vec{p}_B$ respectively and the $x$-axis. Similarly we have
\begin{equation} 
P_{k}\left(1 - \frac{{\theta}^2p_{x_1}^2}{2}\right) P_{k}\left(1 -
\frac{{\theta}^2p^2_{x_2}}{2}\right) P_{k}\left(1 -
\frac{{\theta}^2p^2_{x_3}}{2}\right) = \frac{1}{(2{\pi})^3} \int
d{\alpha}_1 d{\alpha}_2 d{\alpha}_3
e^{iR{\theta}\vec{p}_{k}{\wedge}(\vec{p}_{x_1} + \vec{p}_{x_2} +
\vec{p}_{x_3})}, 
\label{Int3}
\end{equation} 
where now the angles ${\alpha}_i$'s are the angles between the vectors
$\vec{p}_{x_i}$'s and the $x-$axis. Since $R$ is large, the integrals
(\ref{Int0}) with $(A,B)=(k_1,k_2),(k_3,k_5)$ and $(k_4,k_6)$ are
dominated by those values of $\vec{p}_{x_1}$, $\vec{p}_{x_2}$ and
$\vec{p}_{x_3}$ such that $\vec{p}_{x_1} = \vec{p}_{k_4} +
\vec{p}_{k_6}$, $\vec{p}_{x_2} = \vec{p}_{k_1} + \vec{p}_{k_2}$ and
$\vec{p}_{x_3} = \vec{p}_{k_3} + \vec{p}_{k_5}$ respectively, and
correspondingly the integral (\ref{Int3}) is dominated by
$\vec{p}_{x_1} + \vec{p}_{x_2} + \vec{p}_{x_3} = \vec{p}_{k_4} +
\vec{p}_{k_6}$. This is clearly a valid approximation because the
conservation law $\vec{p}_{k_1} + \vec{p}_{k_2} + \vec{p}_{k_3} +
\vec{p}_{k_5} = 0$ is expected to hold (as we explain below) and
because of the large factor of $R$ appearing in the different phases
in (\ref{Int0}) and (\ref{Int3}). After we apply the conservation law
we may reinterpret the angles (say) ${\alpha}_1$ and ${\alpha}_2$ as
the angles made by $\vec{p}_{k_4}$ and $\vec{p}_{k_6}$ and the x-axis
respectively. Using all these ingredients one can convince ourselves
that the sum over $k$ in (\ref{4335}) behaves in the limit as
\begin{eqnarray} 
{\cal S}(46;1235){\simeq}\frac{1}{2l+1} \left\{
\begin{array}{ccc}
k_4&l & l\\
k_6&l&l
\end{array}
\right\}^2 \left\{
\begin{array}{ccc}
k_1&l  & l\\
k_2&l&l
\end{array}
\right\} \left\{
\begin{array}{ccc}
k_3&l  & l\\
k_5&l&l
\end{array}
\right\}.\nonumber
\end{eqnarray} 
We now proceed to the task of rewriting this sum in terms of the
noncommutative plane variables. To this end we use the representation
(\ref{well}) in the form
\begin{eqnarray}
P_{k_4}^2\left(1-\frac{{\theta}^2p_{k_6}^2}{2}\right) &=& \int
\frac{d{\phi}_{k_4}}{2{\pi}} \cos(R{\theta}
\sin{\phi}_{k_4}{p}_{k_4}{p}_{k_6}) \int \frac{d{\phi}_{k_6}}{2{\pi}}
\cos(R{\theta} \sin{\phi}_{k_6}{p}_{k_4}{p}_{k_6})\nonumber\\
&{\simeq}& \int \frac{d{\phi}_{k_4}}{2{\pi}} \frac{d{\phi}_{k_6}}{2{\pi}}
\cos^2(R{\theta} \vec{p}_{k_4}{\wedge}\vec{p}_{k_6}),\nonumber
\end{eqnarray}
where we have used the large $R$ limit to go to the last line,
i.e. since the angles $\phi_4=\phi_6{\simeq}0$ dominate the integrals
in the limit, the two cosines become essentially equal. We have also
reinforced explicitly the symmetry of (\ref{4335}) under the exchange
$k_4{\leftrightarrow}k_6$ on each $6j$-symbol in ${\cal S}$ above (as
is also the case in (\ref{4335})). The ${\phi}_{k_4}$ and
${\phi}_{k_6}$ have the natural interpretation of angles between the
vectors $\vec{p}_{k_4}$ and $\vec{p}_{k_6}$ respectively and the
$x$-axis of the plane. For the case $(A,B)=(k_1,k_2)$, we can use
\begin{eqnarray} 
P_{k_1}\left(1 - \frac{{\theta}^2p_{k_2}^2}{2}\right) = \int
\frac{d{\phi}}{2{\pi}} e^{iR{\theta} \cos{\phi}{p}_{k_1}{p}_{k_2}}. \nonumber
\end{eqnarray} 
However, here ${\phi}$ cannot be interpreted as the angle
between $\vec{p}_{{k}_1}$ (or $\vec{p}_{{k}_2}$) with any
specific axis, but if $\phi_{12}$ is the angle between the two
vectors $\vec{p}_{{k}_1}$ and $\vec{p}_{{k}_2}$ then we can define
$x = {\phi} + {\phi}_{12}$, and write
\begin{eqnarray} 
P_{k_1}\left(1 - \frac{{\theta}^2p_{k_2}^2}{2}\right) =
\int_{{\phi}_{12}}^{2{\pi} + {\phi}_{12}} \frac{dx}{2{\pi}}
e^{-iR{\theta} \sin x \vec{p}_{k_1}{\wedge}\vec{p}_{k_2}}
e^{iR{\theta} \cos x\vec{p}_{k_1} \vec{p}_{k_2}} {\simeq}
e^{-iR{\theta} p_{k_1}{\wedge}p_{k_2}}.\nonumber
\end{eqnarray} 
As before, since $R$ is large, the integral is dominated by the value
$\cos x=0$ or $x=\frac{\pi}{2}$. We can then evaluate the above sum
${\cal S}$ explicitly and find
\begin{eqnarray} 
{\cal S}(46;1235) = \frac{1}{(2l+1)^5}\int
\frac{d{\phi}_{k_4}}{2{\pi}} \frac{d{\phi}_{k_6}}{2{\pi}}
\cos(R{\theta} \vec{p}_{k_1}{\wedge} \vec{p}_{k_2}) \cos(R{\theta}
\vec{p}_{k_3} {\wedge} \vec{p}_{k_5}) \cos^2(R{\theta} \vec{p}_{k_4}
{\wedge} \vec{p}_{k_6}),\nonumber
\end{eqnarray} 
where the symmetry of (\ref{4335}) under the exchanges
$k_1{\leftrightarrow}k_2$ and $k_3{\leftrightarrow}k_5$ is now
explicit. This is essentially the phase of the planar $4$-point function
found in \cite{arefeva}. In order to see this fact more clearly, we
first show that (\ref{4335}) takes now the form
\begin{eqnarray}
{\nu}_1^{(1)} = {\nu}_1^{(4)} &{\simeq}& \frac{a_k(1235)}{R^4} \int
\frac{d{\phi}_{k_4}}{2{\pi}} \frac{d{\phi}_{k_6}}{2{\pi}}
\cos(R{\theta} \vec{p}_{k_1} {\wedge} \vec{p}_{k_2}) \cos(R{\theta}
\vec{p}_{k_3} {\wedge} \vec{p}_{k_5}) \cos^2(R{\theta} \vec{p}_{k_4}
{\wedge} \vec{p}_{k_6}) \nonumber\\
&{\times}& {\delta}^2(\vec{p}_{k_1} + \vec{p}_{k_2} + \vec{p}_{k_3} +
\vec{p}_{k_5}) {\delta}^2 (\vec{p}_{k_1} + \vec{p}_{k_2} +
\vec{p}_{k_4} + \vec{p}_{k_6}),\nonumber
\end{eqnarray}
where we have also made the following interpretation of the
limiting form of the $2$-dimensional fuzzy delta function
\begin{eqnarray} 
(-1)^\frac{m}{2} \frac{R^2}{2l+1} {\delta}_{k,k_0} {\delta}_{m,-m_0}
{\rightarrow} {\delta}(\vec{p}_k + \vec{p}_{k_0}).
\label{nice}
\end{eqnarray} 
The factor $(-1)^{\frac{m}{2}}$ is motivated by (\ref{propagator}),
the factor $2l+1$ is needed in order for (\ref{nice}) to diverge
correctly (in the limit) when $k=k_0$ and $m=-m_0$, while the $R^2$
factor is to restore the correct mass dimension for the delta
function. An identical formula will of course hold for the other
${\mathbb R}^2$ factor, i.e.
\begin{eqnarray}
{\nu}_2^{(1)} = {\nu}_2^{(4)} &{\simeq}& \frac{a_p(1235)}{R^4} \int
\frac{d{\phi}_{p_4}}{2{\pi}} \frac{d{\phi}_{p_6}}{2{\pi}}
\cos(R{\theta} \vec{p}_{p_1} {\wedge} \vec{p}_{p_2}) \cos(R{\theta}
\vec{p}_{p_3} {\wedge} \vec{p}_{p_5}) \cos^2(R{\theta} \vec{p}_{p_4}
{\wedge} \vec{p}_{p_6}) \nonumber\\
&{\times}& {\delta}^2 (\vec{p}_{p_1} + \vec{p}_{p_2} + \vec{p}_{p_3} +
\vec{p}_{p_5}) {\delta}^2 (\vec{p}_{p_1} + \vec{p}_{p_2} +
\vec{p}_{p_4} + \vec{p}_{p_6}). \nonumber
\end{eqnarray}
By putting the above functions ${\nu}_1^{(1,4)}$ and ${\nu}_2^{(1,4)}$
in equation (\ref{1002}), we easily obtain the $4$-dimensional
one-loop planar contributions ${\delta}{\lambda}_4^{(1)}$ and
${\delta}{\lambda}_4^{(4)}$ and consequently the planar contribution
to the $4$-point
function ${\delta}{\lambda}_4^P$. Indeed we have
\begin{eqnarray}
{\delta}{\lambda}_4^{(1)}(1235) = {\delta}{\lambda}_4^{(4)}(1235) &=&
\frac{a(1235)}{R^4{\pi}^4} {\delta}^4 (\vec{p}_{1} + \vec{p}_{2} +
\vec{p}_{3} + \vec{p}_{5}) \nonumber\\
&{\times}& \int_{\Lambda}d^4 \vec{p}_4 \frac{\cos(R{\theta}
\vec{p}_{1} {\wedge} \vec{p}_{2}) \cos(R{\theta} \vec{p}_{3} {\wedge}
\vec{p}_{5}) \cos^2(R{\theta} \vec{p}_{4} {\wedge} \vec{p}_{6})}
{(\vec{p}_4^2 + \frac{m^2}{l})((\vec{p}_1 + \vec{p_2} + \vec{p}_4)^2 +
\frac{m^2}{l})}, \nonumber
\end{eqnarray}
where the notation (in the metric ${\mathbb R}^2 \times {\mathbb
R}^2$) is $p_4^2 = p_{k_4}^2 + p_{p_4}^2, d^4p_4 = \frac{1}{4}
dp_{k_4}^2 dp_{p_4}^2 d{\phi}_{k_4} d{\phi}_{p_4},
{\delta}^4(\vec{p}_1 + \vec{p}_2 + \vec{p}_3 + \vec{p}_5) =
{\delta}^2(\vec{p}_{k_1} + \vec{p}_{k_2} + \vec{p}_{k_3} +
\vec{p}_{k_5}){\delta}^2(\vec{p}_{p_1} + \vec{p}_{p_2} + \vec{p}_{p_3}
+ \vec{p}_{p_5})$ and $\vec{p}_1 {\wedge} \vec{p}_2 = \vec{p}_{k_1}
{\wedge} \vec{p}_{k_2} + \vec{p}_{p_1} {\wedge} \vec{p}_{p_2}$ and
$a(1235) = a_k(1235)a_p(1235)$. 

The associated effective action in
this case can now easily be computed and we find the final result ( with some minor change of notation , namely we denote now the internel momentum $p_4$  as $k$ and denote the externel momentum $p_5$ as $p_4$ )
\begin{equation} 
\frac{g^2_4}{4!} \int_{} \frac{d^4 \vec{p}_1}{(2{\pi})^4}
\frac{d^4 \vec{p}_2}{(2{\pi})^4} \frac{d^4 \vec{p}_3}{(2{\pi})^4}
\frac{d^4 \vec{p}_4}{(2{\pi})^4} {\delta} {\lambda}_4^{P}(1234)
{\delta}^4(\vec{p}_1 + \vec{p}_2 + \vec{p}_3 + \vec{p}_4)
\hat{\phi}_2(\vec{p}_1) \hat{\phi}_2 (\vec{p}_2) \hat{\phi}_2(\vec{p}_3)
\hat{\phi}_2(\vec{p}_4),
\label{finaleffe0}
\end{equation} 
where
\begin{equation} 
{\delta} {\lambda}_4^P(1234) = \frac{32}{3} g^2_4 \int_{}
\frac{d^4\vec{k}}{(2{\pi})^4} \frac{\cos({\theta}^2 \vec{p}_{1}
{\wedge} \vec{p}_{2}) \cos({\theta}^2 \vec{p}_{3} {\wedge}
\vec{p}_{4}) \cos^2({\theta}^2 \vec{k} {\wedge} \vec{P})}
{(\vec{k}^2 + m^2)((\vec{P}+ \vec{k})^2 + m^2)}~,~\vec{P}=\vec{p}_1+\vec{p}_2.
\label{finaleffe}
\end{equation} 
We have employed in above the same definitions as those of the
$2$-point function used in (\ref{ncr4action}), namely $g_4^2=8{\pi}^2{\lambda}_4$ and
$\hat{\phi}_{2}(\vec{p}_1) = 4{\pi}\sqrt{\frac{2}{l^3}}
\hat{\phi}_{NC}(\frac{\vec{p}_1}{\sqrt{l}})$. However, the
noncommutative field $\hat{\phi}_{NC}(\vec{p}_1)$ is now reinterpreted
such that we have $\hat{\phi}_{NC}(\vec{p}_1) {\equiv}
\hat{\phi}_{NC}^{p_{k_1} p_{p_1} {\phi}_{k_1} {\phi}_{p_1}} =
R^4{\phi}^{k_1 p_1 m_1 n_1} \sqrt{(2k_1+1)(2p_1+1)}$ or, in other
words, $\hat{\phi}_{NC}(\vec{p}_1) = {\phi}_{NC}(\vec{p}_1)
\sqrt{(2k_1+1)(2p_1+1)}$.  We notice immediately that equation (\ref{finaleffe}) is exactly the result of \cite{arefeva} upto a numerical factor . More precisely (\ref{finaleffe}) is to be compared with the first term in the expansion of equation $(5)$  of reference \cite{arefeva} which corresponds to the planar contribution to the $4-$point function .

\section{Conclusion}
We have investigated in some detail the problem of obtaining theories
on noncommutative ${\mathbb R}^4$ starting from finite matrix models
defined on $S_F^2 \times S_F^2$. Particular attention was paid to a new
limit that gives a theory on noncommutative ${\mathbb R}^4$ with a UV
cut-off proportional to the inverse of the noncommutativity parameter
$\theta$, and without any mixing between UV and IR degrees of
freedom. 

The new scaling is implemented via the introduction of an intermediate
scale $[2 \sqrt{l}]$. Intuitively, this intermediate scale carries
information about the transition between commutative and
noncommutative regimes of the theory: if we only use modes with
momenta much smaller than this intermediate scale, the theory becomes
commutative, whereas modes with momenta much larger take us the the
noncommutative regime. 

It would be interesting to extend this analysis to theories on $S_F^2$
and $S_F^2 \times S_F^2$ that have fermionic and gauge \cite{iktw} degrees of
freedom, as well as supersymmetric theories \cite{bakuro}. We also see
no obstacle to using this method to study theories that are obtained
from Kaluza-Klein reduction on fuzzy $S^4$ \cite{denjoe}, as well as 
gauge theory on fuzzy $CP^2$ \cite{kitazawa}. 

{\bf Acknowledgments:} It is a pleasure to thank A. P. Balachandran,Denjoe O'Connor and
Peter Pre\v{s}najder for numerous discussions. SV
would like to thank DIAS for warm hospitality during the final stage
of this project. The work of SV is supported in part by DOE grant
DE-FG03-91ER40674.

\bibliographystyle{unsrt}



\end{document}